\documentclass[a4paper,usenatbib]{mnras}
\usepackage{newtxtext,newtxmath}
\usepackage{cancel}
\usepackage[T1]{fontenc}
\usepackage{ae,aecompl}
\usepackage{amsmath} 
\usepackage{graphicx} 
\usepackage[utf8]{inputenc}
\usepackage{qtree}
\usepackage[normalem]{ulem}
\usepackage{gensymb}
\usepackage{xcolor}

\newlength{\figsize}
\newlength{\figdblsize}
\newenvironment{shortitem}
{\begin{list}{$\bullet$}{\topsep=0pt\itemsep=0pt\parsep=0pt\parskip=0pt\leftmargin=12pt}}
{\end{list}}
\newcommand{\bsi}{\begin{shortitem}}
\newcommand{\esi}{\end{shortitem}}
\newenvironment{shortsubitem}
{\begin{list}{$\circ$}{\topsep=0pt\itemsep=0pt\parsep=0pt\parskip=0pt\leftmargin=12pt}}
{\end{list}}
\newcommand{\bssi}{\begin{shortsubitem}}
\newcommand{\essi}{\end{shortsubitem}}

\newcommand{\eagle}{\mbox{\sc Eagle}}
\newcommand{\subfind}{\mbox{\sc Subfind}}
\newcommand{\Jstar}{\mbox{$\vec{J}_{\star}$}}

\newcommand{\Msun}{\hbox{$\rm\,M_{\odot}$}}

\newcommand{\Ngc}{\mbox{$N_\mathrm{gc}$}}
\newcommand{\nside}{\mbox{$n_\mathrm{side}$}}

\newcommand{\Dt}{\mbox{$\Delta \theta$}}
\newcommand{\DTTj}{\mbox{$D/T_{\vec{\mathrm{J}}_{\mathrm{b}}=0}$}}
\newcommand{\DTTe}{\mbox{$D/T_{\mathrm{\epsilon>0.7}}$}}
\newcommand{\DTTdt}{\mbox{$D/T_{\mathrm{\Dt<30\degree}}$}}
\newcommand{\DTTcr}{\mbox{$D/T_{\mathrm{CR}}$}}

\newcommand{\Fig}[1]{Fig.~\ref{fig:#1}}

\newcommand{\Eq}[1]{Equation~\ref{eq:#1}}
\newcommand{\Sec}[1]{Section~\ref{sec:#1}}
\newcommand{\App}[1]{Appendix~\ref{app:#1}}

\title[Identifying kinematically distinct galactic components]{Using angular momentum maps to detect kinematically distinct galactic components}
\author[Irodotou \& Thomas]
{{Dimitrios Irodotou$^1$}\thanks{E-mail: Dimitrios.Irodotou@sussex.ac.uk},
 {Peter A.~Thomas$^1$} \\
 {}$^1$Astronomy Centre, University of Sussex, Falmer, Brighton BN1 9QH, UK}

\begin{document}

\pagerange{\pageref{firstpage}--\pageref{lastpage}} \pubyear{2020}

\maketitle

\label{firstpage}
\begin{abstract}
  In this work we introduce a physically motivated method of performing disc/spheroid decomposition of simulated galaxies, which we apply to the \eagle\ sample. We make use of  the HEALPix
  package to create Mollweide projections of the angular momentum map of each galaxy's stellar particles. A number of features arise on the angular momentum space which allows us to decompose galaxies and classify them into different morphological types. We assign stellar particles with angular separation of less/greater than 30 degrees from the densest grid cell on the angular momentum sphere to the disc/spheroid components, respectively.  We analyse the spatial distribution for a subsample of galaxies and show that the surface density profiles of the disc and spheroid closely follow an exponential and a Sersic profile, respectively.  In addition discs rotate faster, have smaller velocity dispersions, are younger and are more metal rich than spheroids.  Thus our morphological classification reproduces the observed properties of such systems. Finally, we demonstrate that our method is able to identify a significant population of galaxies with counter-rotating discs and provide a more realistic classification of such systems compared to previous methods. 
\end{abstract}


\section{Introduction} \label{sec:Intro}
Exploring how and when galactic components form is an essential step towards understanding the formation and evolution of galaxies. Hence, a method that not only accurately identifies the constituent stellar populations but provides an additional way of exploring their dynamics is of great importance. The plethora of methods as presented below shows the usefulness of decomposing either observed or simulated galaxies, and the purpose of this work is to introduce a pioneering method of detecting kinematically distinct components and exploring their properties.

Photometric decompositions have been long used for splitting distinct stellar populations. A central component (``bulge'') is separated from an extended one (``disc'') with the use of different photometric profiles; however different methods can lead to variations in the contributions from each component \citep{CCC20}. Various software packages \citep[e.g., GIM2D, GALFIT, BUDDA and IMFIT, see][ respectively]{SWV02,PHI02,SGA04,E15} allow the fit of a single- or double-component models and provide a plethora of profiles (e.g., Nuker law, Sérsic (de Vaucouleurs) profile, exponential, Gaussian or Moffat/Lorentzin functions). This allows some of these codes to perform multi-component decomposition and identify, in addition to a disc and a bulge; nuclear rings, lenses and bars.

Many methods of bulge/disc decomposition in simulated galaxies have been proposed. One method assumes that the bulge has zero net angular momentum, hence estimates the disc-to-total (D/T) ratio by assigning to the bulge the sum of the mass of the counter rotating particles multiplied by 2. This method has been extensively used \citep[e.g.,][]{CMF10,CSF18,TFT19} even though it contains a crude assumption regarding bulge kinematics. A different method \citep{ANS03} follows a kinematic decomposition in order to estimate the D/T ratio of simulated galaxies by analysing the circularity of the orbits of stellar particles. This is defined (for a given stellar particle) as the ratio between the component of the angular momentum which is normal to the rotation plane and the maximum angular momentum for a stellar particle with the same binding energy (i.e., if on a circular orbit in the rotation plane). Many authors combined this idea with additional (spatial and/or binding energy) criteria in order to assign stellar particles to disc, bulge and inner/outer halo components \citep{TWS12, CPL15, PMM15, MGG19, RTP19}. These
attempts to more accurately identify distinct components unavoidably increase the complexity and free parameters of the decomposition. Finally, \cite{GMG19} introduced a hybrid technique that combines spatial and kinematic criteria in order to study bulges. However, as noted by \cite{JPN20} this method (i.e., the circularity parameter) does not always agree with visual classifications of morphology.

A new era of decomposition software utilises machine-learning techniques to train neural networks to identify the correct profiles for each component \citep[e.g.,][]{DHD18}. For example, a method introduced by \cite{DHB18} classified SDSS morphology based on Convolution Neural Networks, while \cite{DHD02} introduced a machine learning algorithm to identify kinematic structures in IllustrisTNG galaxies.

Lastly, citizen based projects like Galaxy Zoo \citep{MG20} follow a different path by allowing volunteers to perform the decomposition \citep{LMK20}.

The aim of this paper is to present a new decomposition method and compare galactic and component properties with local ($z\sim0.1$) observational data. Our method not only provides a physically-motivated spheroid/disc decomposition framework, but is also able to identify (in some cases multiple) kinematically distinct components based on their detailed representation on the angular momentum sphere. This gives one the unique ability of visually inspecting the angular momentum space of galaxies and utilising this information to study the imprint of secular \citep{KK04} and violent \citep{T77} processes on the constituent stellar components.

This paper is organised as follows. In \Sec{Methodology} we describe our sample and introduce our method. In \Sec{Classification} we present a sample of angular momentum maps drawn from the \eagle\ simulation and characterise the main morphological types. A detailed analysis of our spheroid-disc decomposition is presented in \Sec{Results}, a discussion regarding counter-rotating discs is presented in \Sec{Discussion} and our conclusions are summarised in \Sec{Conclusions}.

\section{Methodology} \label{sec:Methodology}
In \Sec{method:The galaxy sample} we describe how we select galaxies for our study, and in \Sec{method:Decomposition} we describe the decomposition into spheroid and disc components on the basis of the angular momentum distribution of the stellar particles.

\subsection{The galaxy sample} \label{sec:method:The galaxy sample}
We apply our method to the RefL0100N1504  flavour of the \eagle\ simulation \citep{SCB15,CSB15}. We define galaxies as gravitationally bound sub-structures within FoF structures \citep[i.e., they consist of particles that share the same subgroup and group number][]{TET17}, where the former are identified by the \subfind\ algorithm \citep{SWT01}. We focus on galaxies with stellar masses $M_{30} > 5\times 10^9\Msun$ where $M_{30}$ represents the stellar mass within a 30\,kpc spherical aperture and we exclude all particles with separation more than 30\,kpc from the galactic centre (defined as the position of the most bound particle). This guarantees that even the least massive galaxy is resolved with more than 3\,000 stellar particles. Although our method can be used for lower particle numbers, it will become less reliable as the particle number significantly drops below 3\,000, due to numerical effects which will lead to scattering of orbits.

\subsection{Decomposition} \label{sec:method:Decomposition}
The HEALPix\footnote{https://healpix.sourceforge.io} sphere \citep{GWH99,GBH02} is hierarchically tessellated into curvilinear quadrilaterals where the area of all grid cells at a given resolution is identical. We set $\nside=2^4$ which is a parameter that represents the resolution of the grid\footnote{We performed tests for $\nside=2^i$ where $i$=4, 5, 6 and there is no significant change in the D/T ratios and the results presented in this work.} (i.e., the number of divisions along the side of a base-resolution pixel). This results in $3\,072$ HEALPix grid cells ($\Ngc=12\nside^2$) available over the sky which roughly corresponds to the same number of stellar particles our least massive galaxy has. In practice, our decomposition method follows the steps
\begin{enumerate}
\item We define the angular momentum vector of a stellar particle $i$ as
\begin{align} \label{eq:j_i}
\vec{j}_i = m_i (\vec{r}_i - \vec{r}_\mathrm{mb}) \times (\vec{v}_i - \vec{v}_\mathrm{CoM})\; ,
\end{align}
where $m_i$ is its mass, $\vec{r}_i$ is its position vector, $\vec{r}_\mathrm{mb}$ is the position vector of the most bound particle (defined by \subfind), $\vec{v}_i$ is the velocity vector of the particle and $\vec{v}_\mathrm{CoM}$ is the velocity vector of the centre of mass defined as the collection of all stellar, gas, black hole and dark matter particles that belong to same galaxy as particle $i$ and are within a 30kpc spherical aperture centred on the potential minimum. We define the total stellar angular momentum vector as
\begin{align} \label{eq:j_star}
\Jstar = \sum_{i=1}^N\vec{j}_i\; ,
\end{align}
where the summation goes over all stellar particles belonging to the corresponding galaxy.

\item We convert the angular momentum unit vector of all stellar particles from Cartesian to spherical coordinates (we use $\alpha \in [-180\degree,180\degree]$ as the azimuth angle and $\delta \in [-90\degree,90\degree]$ as the elevation angle) which we provide to HEALPix in order to generate the pixelisation of the angular momentum map. 

\item We smooth the angular momentum map with a top-hat filter of angular radius 30$\degree$ and then identify the densest grid cell (i.e., the coordinates $(\alpha_\mathrm{den},\,\delta_\mathrm{den})$ of the grid cell on the angular momentum sphere that contains most stellar particles).

\item We calculate the angular separation of each stellar particle from the centre of the densest grid cell as
\begin{align} \label{eq:delta_theta}
\Dt_{i} = \arccos\big(\sin\delta_\mathrm{den}\sin\delta + \cos\delta_\mathrm{den}\cos\delta\cos(\alpha_\mathrm{den} - \alpha)\big)\; ,
\end{align}
where the index $i$ goes over all stellar particles belonging to the corresponding galaxy.

\item We assign to the disc component all stellar particles that satisfy
\begin{align} \label{eq:criterion}
\Dt_{i} < 30 \degree\; ,
\end{align}
and the remaining particles to the spheroid. Hence, we directly select stellar particles based on their orbital plane. The choice of $30\degree$ was motivated by the visual inspection of the angular momentum maps (see \Sec{Classification}) and the work of \cite{P20} who showed that for an axisymmetric disc with a flat rotation curve the eccentricity of a particle's orbit can be written as
\begin{align} \label{eq:eccentricity}
\epsilon = 0.8 \mathrm{cos}(\theta)\; ,
\end{align}
where $\theta$ is the angle that quantifies the tilt of the orbit with respect to the disc plane. For our $30 \degree$ criterion this results in particles with minimum value of $\epsilon \sim 0.7$, a limit commonly used for disc particles \citep[e.g.,][]{SWS09,MPS14}. In addition, our $30\degree$ criterion results in good agreement and tight correlations with other commonly used morphological parameters (see \Sec{results:Correlations with other methods} for more details).

\item Our method by construction will assign particles to the disc component even for the most idealised dispersion-supported system (i.e., a systems whose particle's angular momenta will be perfectly uniformly distributed on the HEALPix sphere). That is because, even for an isotropic distribution over the sky, there will be a small fraction of particles whose angular momentum lies within 30 degrees of the nominal (directed) rotation axis. The following expression for the D/T ratio accounts for this "artificial" increase of the disc component
\begin{align} \label{eq:Correction}
D/T_{\mathrm{\Dt<30\degree}}  = \frac{1}{1-\chi}\left( \frac{N_\mathrm{\Dt<30\degree}}{N_\mathrm{all\ sky}} - \chi\right)\; ,
\end{align}
where $N_\mathrm{\Dt<30\degree}$ and $N_\mathrm{all\ sky}$ are the number of stellar particles within $30\degree$ of the direction of the rotation axis and the whole sky, respectively, and
\begin{align} \label{eq:Chi}
\mathrm{\chi = \frac{1-cos(30\degree)}{2}}\; .
\end{align}
\end{enumerate}

Finally, we note that in this work we prefer not to limit the spatial extent of the spheroid (i.e., not split it into what is usually termed as a stellar halo and a bulge), since as recently discussed by \cite{CSF18,GMG19} there is no physical criterion that determines the boundary between a bulge and a halo. In addition we do not attempt to split the disc into a cold and warm (thin and think) component \citep{OMM18}. However, in a future work we intend to explore the imprint of such components on the angular momentum maps.

\section{Morphological classification} \label{sec:Classification}
In \Sec{classification:Distinct categories} we present angular momentum maps of the 100 most massive central \eagle\ galaxies and classify them into distinct categories, and in \Sec{classification:A representative sample} we study in more detail the maps of an example representative of each category, and compare three different methods of decomposing disc and bulge.

\subsection{Distinct categories} \label{sec:classification:Distinct categories}
\begin{figure*}
\centering \includegraphics[width=\figdblsize]{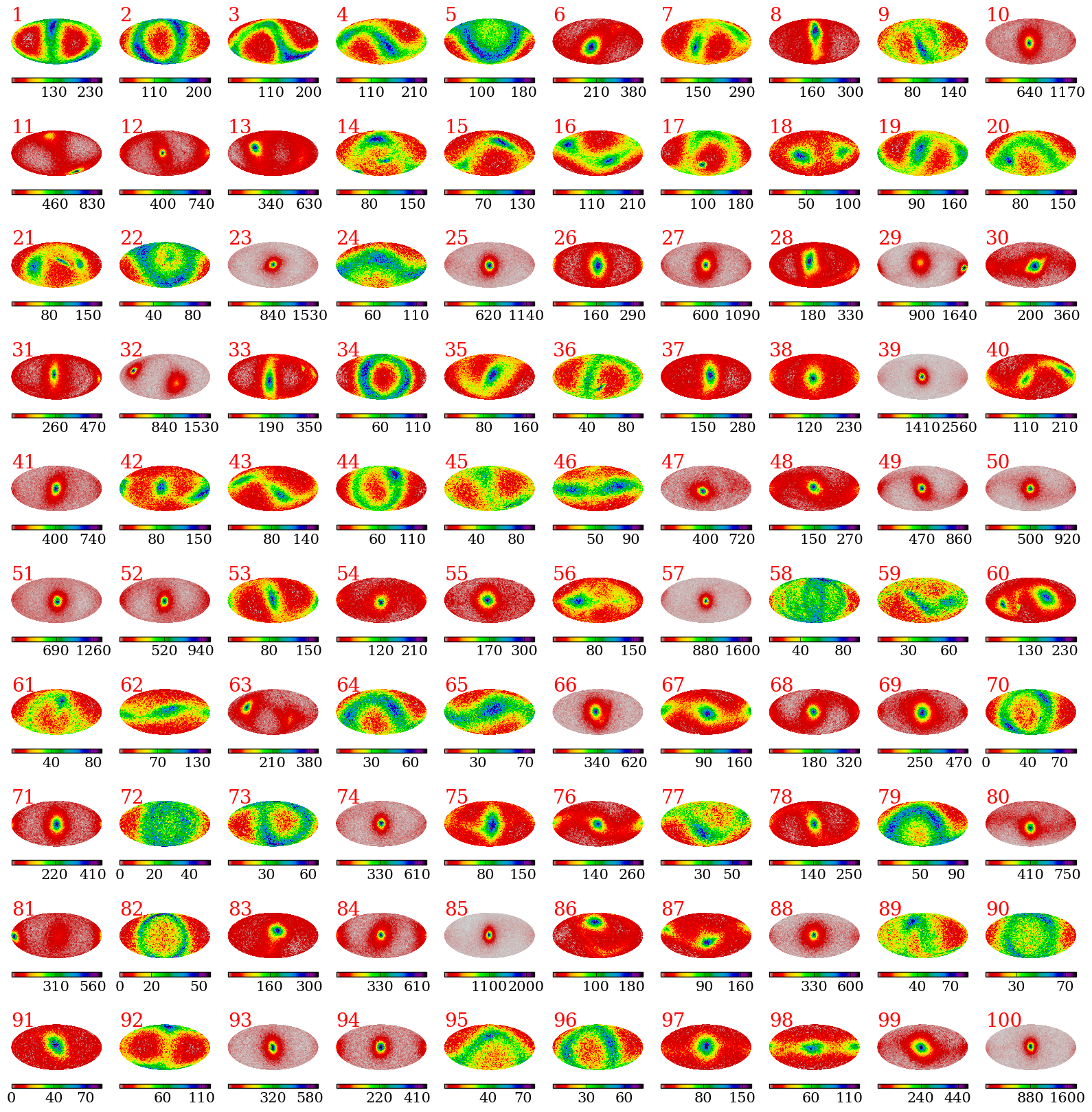}
\caption{A sample of the 100 most massive central galaxies in RefL0100N1504. In each panel the red number on the top left corner refers to the galaxy's group number and the colour bar represents the number of particles per grid cell. Note that for visual purposes the galaxies have been rotated based on the process described in \App{A}.}
\label{fig:SRAEl}
\end{figure*}

\Fig{SRAEl} displays on a Mollweide, equal-area projection the angular momentum maps of the 100 most massive central \citep[see][]{TET17} galaxies. For visual purposes, all particles have been rotated so that \Jstar\ points towards the reader (see \App{A}). Several distinct categories of structure are visible:

\begin{shortitem}
\item {\bf Rotationally-supported systems:} The simplest behaviour is a single cluster of dense grid cells (e.g., galaxy 39, 85, 100), usually closely aligned with the total angular momentum of the galaxy, \Jstar.  These are well-ordered disc-dominated galaxies.   Interestingly, the majority of these (e.g., galaxy 25, 57, 94) also show a slight antipolar excess of counter-rotating particles -- we discuss this further in \Sec{Discussion} below.

\item {\bf Dispersion-supported systems:}
  These do not show a single, dominant density peak in the angular momentum map.  There are a variety of morphologies:
\begin{shortsubitem}
\item {\bf Double-peak systems:} There are several examples of systems with two, distinct density peaks in the angular momentum maps, either relatively isolated (e.g., galaxy 18, 40, 87), or buried within a great circle of orbits (e.g., galaxy 2, 46, 70). For these galaxies there is no well-defined, ordered rotation; hence they represent dispersion-dominated systems. Where two density peaks are visible, these tend to be opposite each other on the sky: in such an orientation the minor axes align and the mixing and precession of orbits in the merger remnant is heavily suppressed. They are likely the result of either a merger of progenitors with opposite angular momentum \citep{B88,KWU15} or the accretion of counter-rotating gas which settles and forms stars with angular momentum in the opposite direction \citep{VPC07,CMP13,ANA14}.

\item {\bf Great circle systems:} Galaxies with a ring (great circle) of high-density points in the map (e.g., galaxy 34, 82, 96); depending on the orientation of the galaxy it can appear as a horizontal S-shaped (e.g., galaxy 3, 19, 65) or as a U (e.g., galaxy 5, 22, 77) or upside down U (e.g., galaxy 20, 64, 79). It is interesting to note that this great-circle structure is much more common than having orbits that are more uniformly distributed (e.g., galaxy 58, 72, 90) across all directions in space. This is anticipated since mergers are not expected to mix orbits up evenly in phase space. Instead, realistic merger remnants have angular momentum distributions that are composed of varying contributions from different orbital families, such as short- and long-axis tube orbits \citep{VM98,DVM11}.  Some of the dispersion-dominated systems (e.g., galaxy 5, 22, 34) have a uniform density of orbits around a great circle, which reflects that for every direction in space there are roughly equal numbers of stellar particles with positive and negative angular momenta; others show a clear density peak (e.g., galaxy 20, 44, 61) which suggests a relatively minor merger.

\end{shortsubitem} 

\item {\bf Multiple-peak (merging) systems:} A very few systems show three or more density peaks (e.g., galaxy 14, 21, 60). These are unlikely to be stable and may represent the early stages of mergers of a third galaxy onto a double-peaked system. It is beyond the scope of this work to study the formation path and evolution of such systems and their components, however we plan to do so in a future work.

\end{shortitem} 

\subsection{Detailed kinematics} \label{sec:classification:A representative sample}
\begin{figure*}
  \centering
  \includegraphics[width=0.95\figsize]{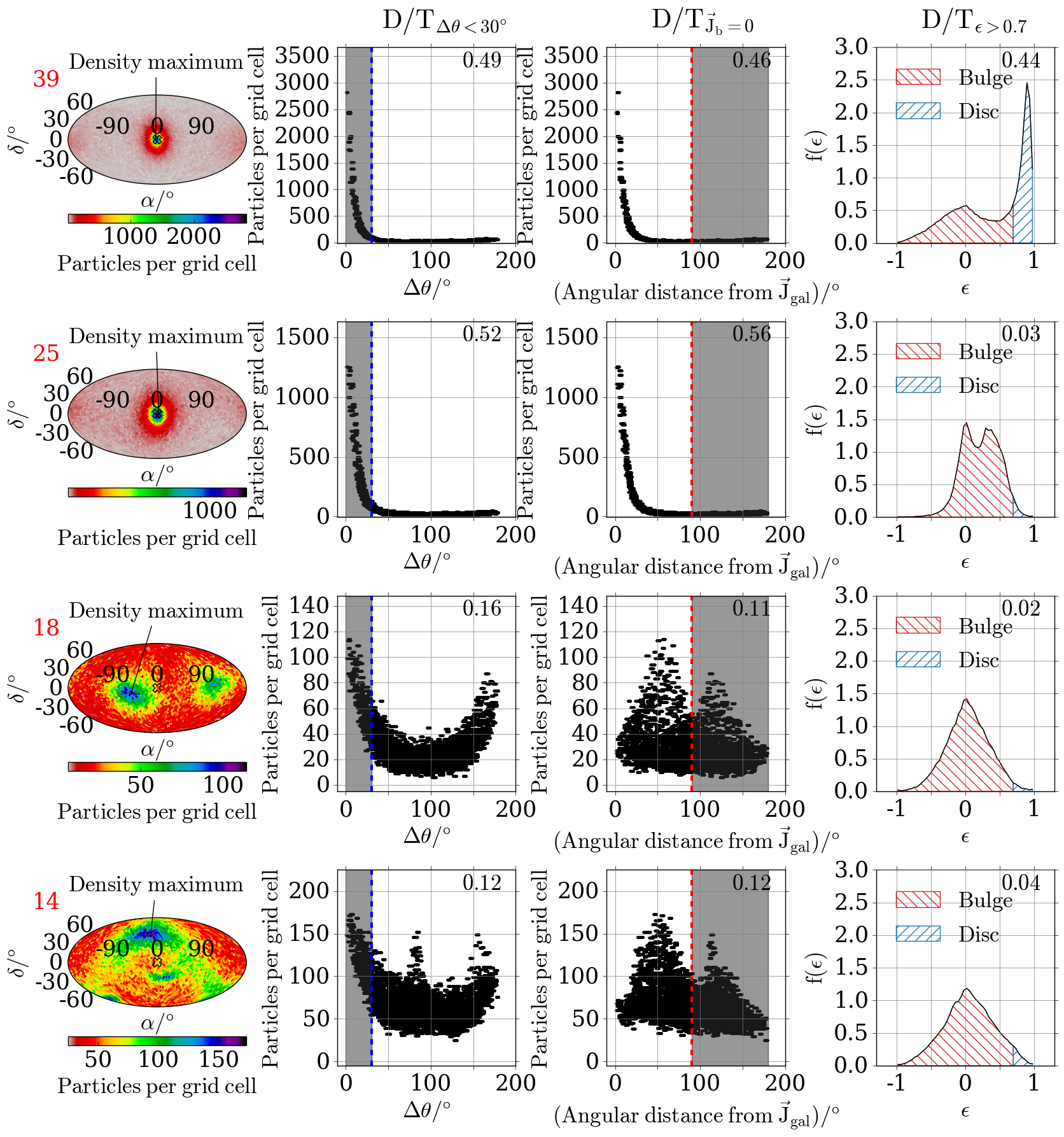}\hfill
  \includegraphics[width=0.95\figsize]{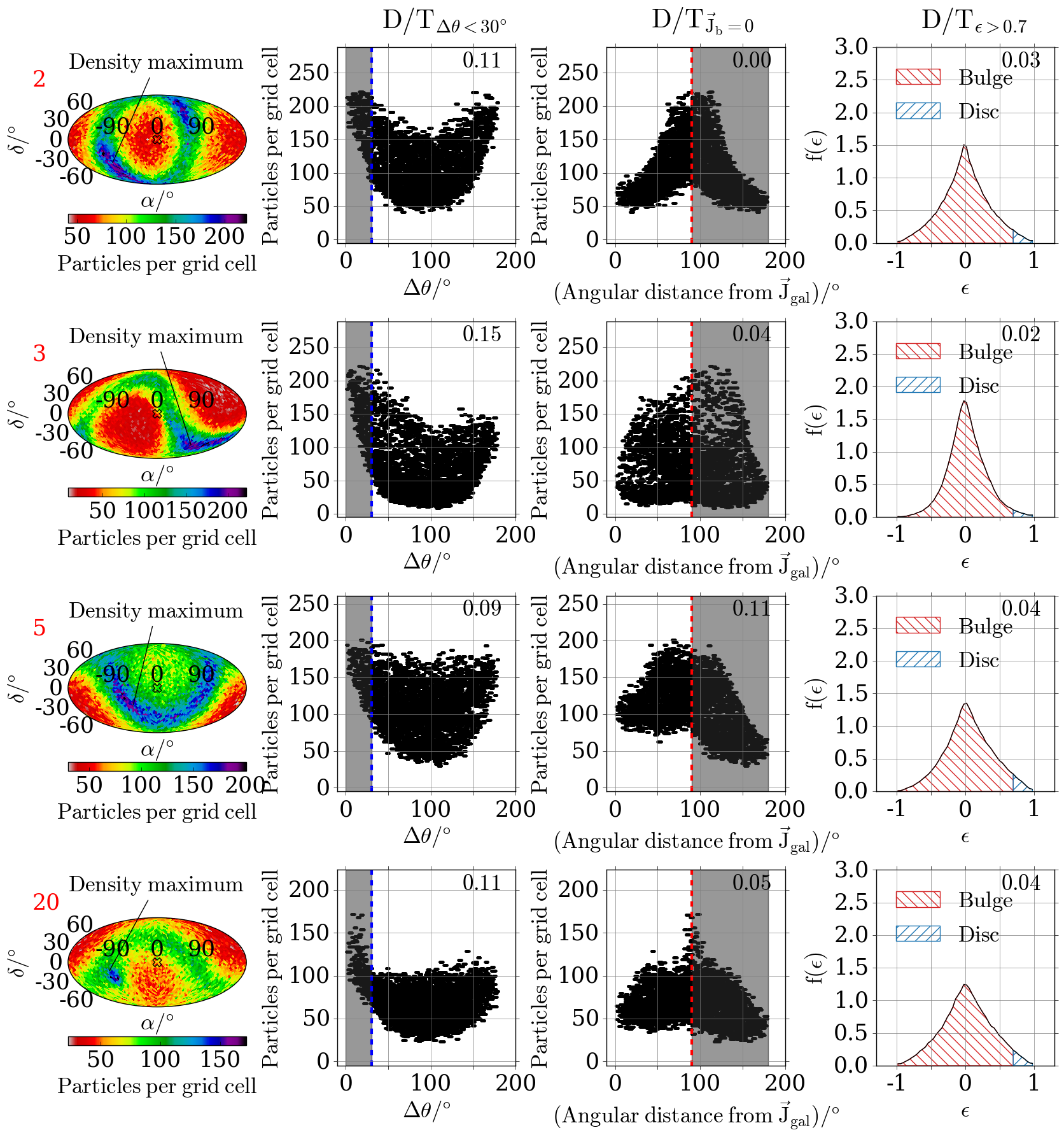}
  \caption{A sample of 8 galaxies drawn from the different morphological classifications described in \Sec{classification:Distinct categories}. Left 4 panels: rotationally-supported (galaxy 39, 25), double- (galaxy 18) and multi-peak (galaxy 14); right 4 panels: dispersion-supported systems.  The first column for each galaxy displays on a Mollweide projection the angular momentum maps, where the red number on the top left corner in each row refers to each galaxy's group number and the black line points to the densest grid cell. As in \Fig{SRAEl} the galaxies have been rotated so that \Jstar\ (hollow X symbol) points towards the reader. The second column contains the number of particles in each grid cell as a function of the angular separation from the densest grid cell (as defined by \Eq{delta_theta}). The blue dashed vertical line marks $\Dt =30\degree$ and the gray shaded region highlights all stellar particles that belong to the disc component (i.e., have $\Dt <30\degree$). The third column shows the number of particles in each grid cell as a function of the angular separation from the angular momentum vector. The red dashed vertical line marks angular distance of $90\degree$
    and the gray shaded region highlights all counter-rotating stellar particles. The fourth column contains the PDF of the stellar mass-weighted distribution of the orbital circularity of its stellar particles. The red and blue hatches represent stellar particles with $\epsilon<0.7$ and $\epsilon>0.7$, respectively to account for a bulge and disc component. The text on the top right corner of the latter three columns represents for each galaxy an estimate of the D/T ratio based on three different methods (see text for more details).}
\label{fig:SMD}
\end{figure*}

\begin{figure*}
  \centering \includegraphics[width=0.95\figsize]{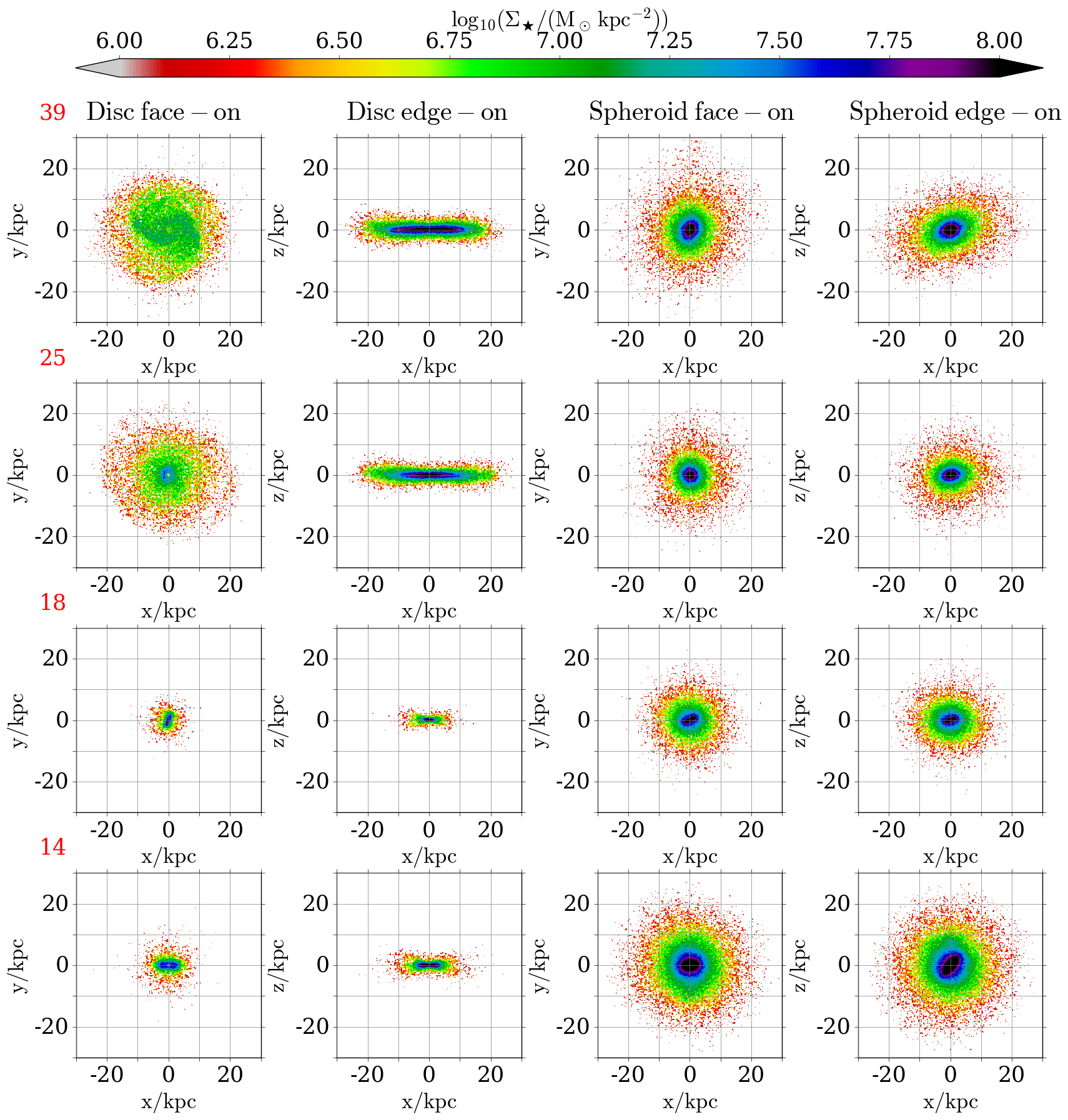}\hfill
  \includegraphics[width=0.95\figsize]{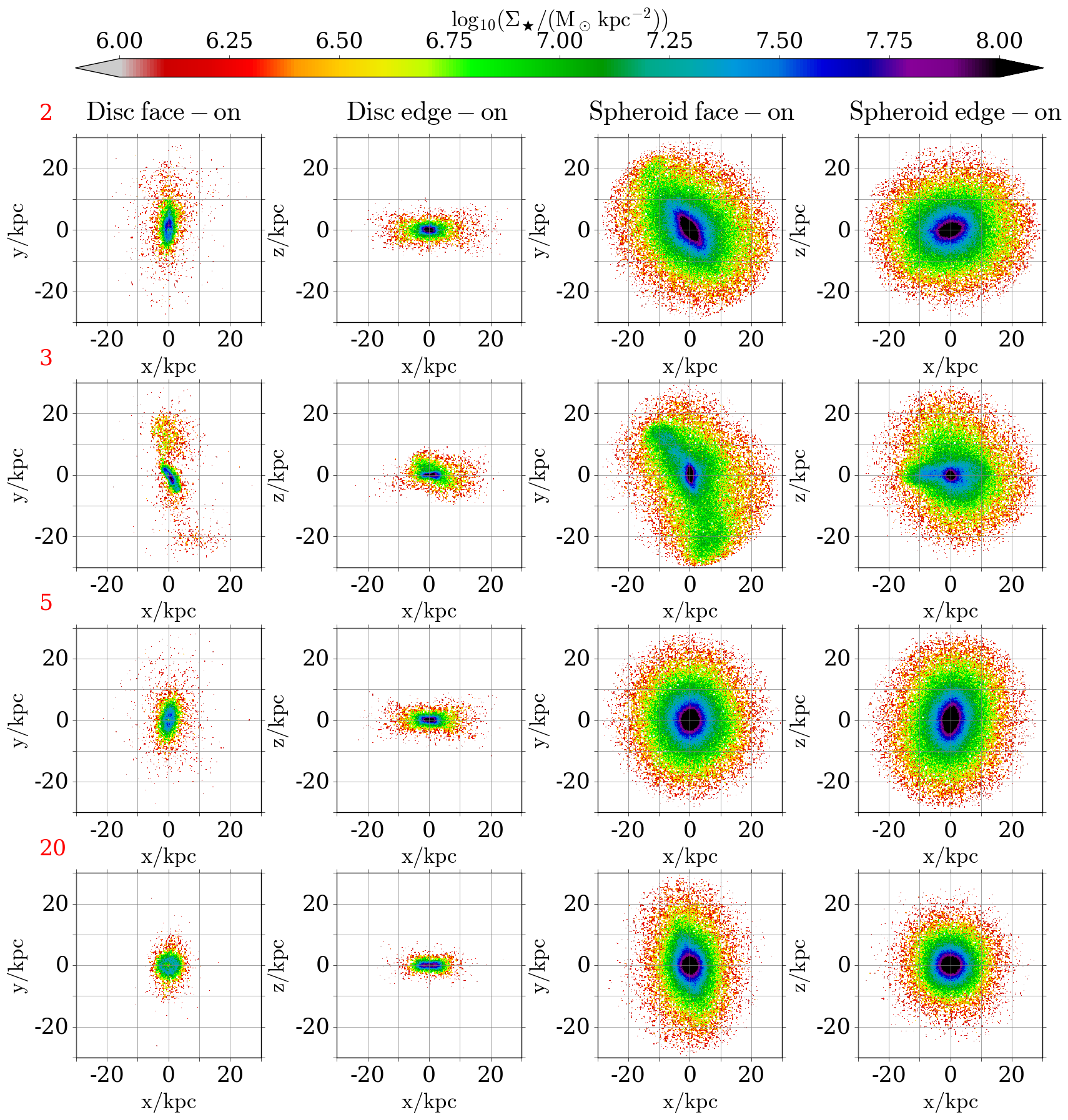}
\caption{The stellar surface density of the components identified by our method for the sample of the 8 galaxies shown in \protect\Fig{SMD}. The red number on the top left corner in each row refers to each galaxy's group number. The face-on and edge-on projections for the disc and spheroid are shown as indicated at top of each column.}
\label{fig:SSD}
\end{figure*}

The first and fith columns of \Fig{SMD} display on a Mollweide projection the angular momentum maps of galaxies representative of each one of the categories and subcategories identified in \Sec{classification:Distinct categories}. In addition, for each galaxy it shows the grid cell density as a function of the angular separation from the densest grid cell (second and sixth columns) and from the angular momentum vector (third and seventh columns), and the distribution of the orbital circularity parameter (fourth and final columns). We follow \cite{TCM19} and defined the latter for a given stellar particle $i$ as
\begin{align} \label{eq:epsilon}
\epsilon_{i} = \frac{L_{\perp,i}}{L_{\perp,\mathrm{max}}(E<E_{i})}\; ,
\end{align}
where $L_{\perp,i}$ is the component of the angular momentum which is perpendicular to the rotation plane and $L_{\perp,\mathrm{max}}(E<E_{i})$ the maximum value of the same component achieved by any stellar particle with binding energy less than that of particle $i$.
  The red and blue hatched regions represent the distribution of the circularity parameter for bulge and disc particles defined as $\epsilon<0.7$ and $\epsilon>0.7$, respectively \citep{GGM17,R20}.  The \DTTdt\ ratio for each galaxy in the second/sixth columns results from the method introduced in this work which assigns to the disc component stellar particles with $\Dt<30\degree$ and the remaining to the spheroid. The \DTTj\ ratio shown for each galaxy in the third/seventh columns follows the assumption that the bulge has zero net angular momentum, hence its mass equals the mass of all counter-rotating particles multiplied by 2, and the remaining stellar particles are assigned to the disc component. Finally, the \DTTe\ ratio produced for each galaxy following \Eq{epsilon} is shown in the fourth/final columns.

\begin{shortitem}
\item Galaxy 39: The majority of stellar particles have angular momenta well aligned both with the densest grid cell (second column) and the galactic angular momentum vector (third column). As discussed in \Sec{classification:Distinct categories} galaxy 39 is a rotationally-supported galaxy since its ordered motion results in a relatively high \DTTdt\ ratio (0.49). This value is quite close to the \DTTj\ and the \DTTe\ ratios (0.46 and 0.44, respectively).

\item Galaxy 25: An interesting example of a disc-dominated galaxy which has orbits in a plane, but they are not circular. This is reflected in its relatively high \DTTdt\ (0.52) and \DTTj\ (0.56) ratios, but a considerably lower \DTTe\ ratio (0.03). Even though a well-ordered component appears as a second peak in the distribution of $\epsilon$, the strict criterion of $\epsilon > 0.7$ fails to include to the disc component stellar particles with non-circular orbits. Hence, galaxy 25 (and of course others similar in nature) would have been misclassified by that method (see also discussion in \Sec{results:Correlations with other methods}). 

\item Galaxy 18: The two density peaks on the angular momentum map are almost counter-rotating with respect to each other as can be seen on the \Dt -density plane (second column). However, the picture is different when the angular separation is calculated from the angular momentum vector (third column) and not from the densest grid cell. Since neither of the structures are aligned with \Jstar, particles from both clusters of dense grid cells will contaminate the counter-rotating (gray shaded) region. This results in an artificial overestimation of the number of counter-rotating particles, hence a lower \DTTj\ and \DTTe\ ratios compared to \DTTdt.  We discuss the effect of including counter-rotating particles in the definition of the disc component in \Sec{Discussion}.

\item Galaxy 14: There is a clear advantage of our method compared to the other ones presented in \Fig{SMD}, which do not have the ability to identify three kinematically distinct components. The three distinct peaks on the \Dt -density plane are not present in the last two panels corresponding to galaxy 14 on the fourth row.

\item Galaxy 2: A perfect example of a balance between equal numbers of rotating and counter-rotating stellar particles.  Galaxies with almost zero net rotation can be the result of either two counter-rotating structures or a uniform distribution of stellar orbits. Hence, in order to understand their formation one first must accurately identify their distinct kinematic components. If the mechanism behind the formation of such galaxy is a merger of two counter-rotating galaxies \citep{HCY08,LSB18a} or accretion of counter-rotating gas, then the remnant disc should include both disc structures instead of none. Both \DTTj\ and \DTTe\ values are quite close to zero since these methods only use information regarding the total angular momentum of the galaxy. The inability of angular momentum based decomposition methods to reveal the true nature of such galaxies has been previously reported by \cite{CSF18}. However, our method is capable of identifying the two counter-rotating structures in opposite parts of the angular momentum sphere (embedded within a great circle), hence providing a more realistic D/T estimate -- see also discussion in \Sec{Discussion} where we investigate the inclusion of counter-rotating particles in our definition of D/T.

\item Galaxies 3, 5, 20: these all represent dispersion-supported system as discussed in \Sec{classification:Distinct categories} and are similar in nature to galaxy 2, but with more asymmetry and a variety of strengths for the main density peak(s).
\end{shortitem}

In summary, all galaxies apart from galaxy 39 and 25 have relatively similar D/T ratios, however the imprint they leave on the angular momentum maps are far from similar. This means that their formation histories and the mechanisms that gave rise to these features were not identical. Identifying these kinematically distinct components gives one the ability to isolate and study them in more detail, in an attempt to understand galaxy formation and evolution, as we briefly do below.

\subsection{Morphology of the individual components} \label{sec:spatial}
\begin{figure*}
\centering \includegraphics[width=\figdblsize]{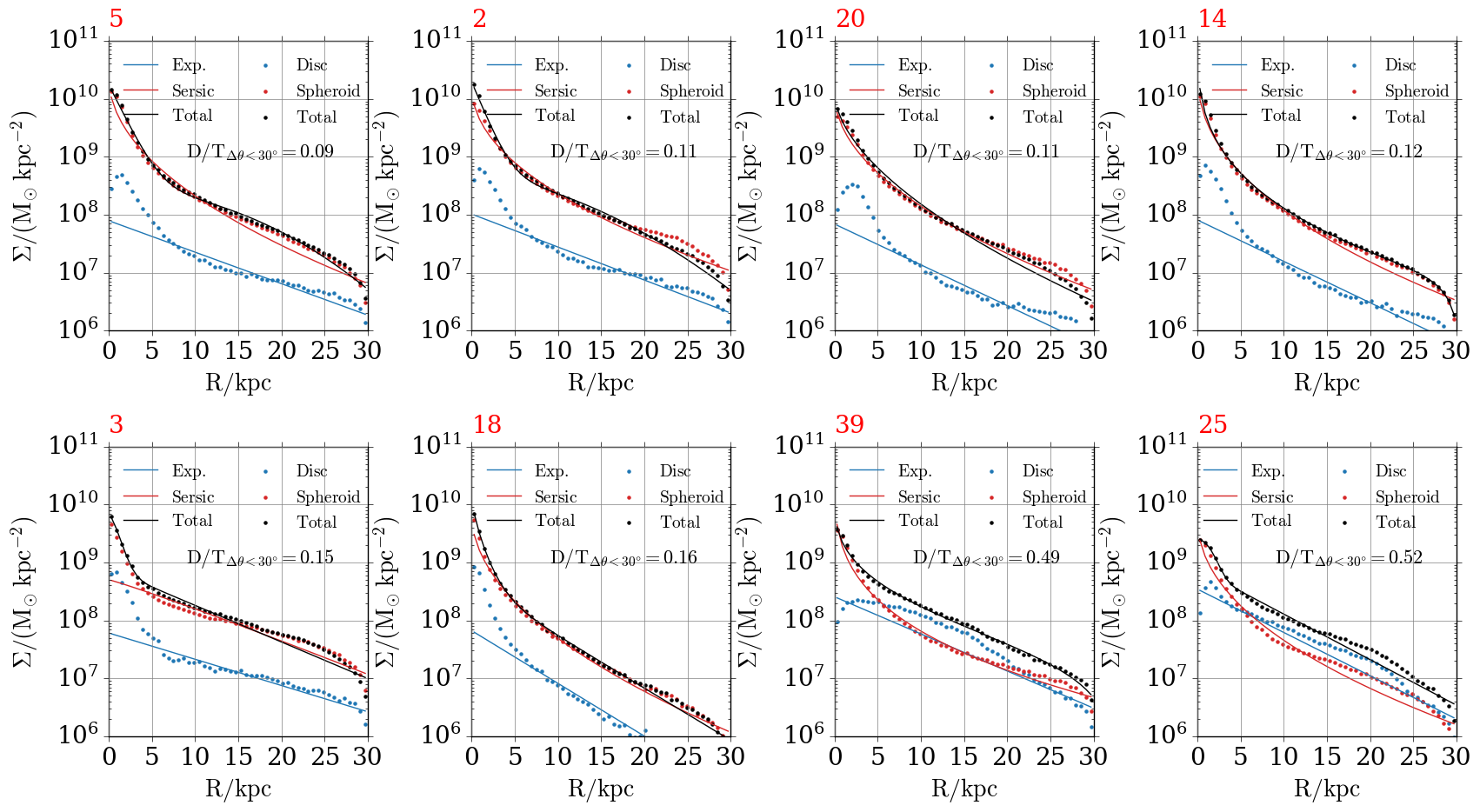}
\caption{Face-on stellar surface density profiles for the sample of the 8 galaxies shown in \protect\Fig{SMD} sorted based on their \DTTdt\ values. These galaxies represent a diverse subsample as indicated by each galaxy's \DTTdt\ ratio. The red number on the top left corner in each panel refers to each galaxy's group number. The black symbols represent the total stellar surface density and the black curve the two-component fit which consists of a Sersic and an exponential profile (see text for more details). The blue and red symbols represent, respectively, the disc and spheroid components' stellar surface densities which are fit with an exponential (blue curve) and a Sersic (red curve) profile, respectively.}
\label{fig:SSDP}
\end{figure*}

\Fig{SSD} shows the face-on and edge-on projection of the disc and spheroid components identified by our method (see \Sec{method:Decomposition}) for the sample of galaxies presented in \Fig{SMD}. Galaxies 39 and 25 which are rotationally-supported galaxies, have prominent spiral arms (face-on projections) and a well-defined thin discs (edge-on projections) akin to grand-design spiral galaxies. Galaxy 18, which showed as a double peak in \Fig{SMD}, is seen to contain a very compact and elongated disc, similar to the dispersion-supported systems. The triple-peak nature of Galaxy 14 shows as a small sub-clump within the spheroid, but Galaxy 3 shows a much stronger asymmetry, suggestive of a more significant merger, but not evident in the kinematic analysis -- that shows the importance of using more than one technique to characterise the structure of galaxies.

Investigating further the morphology of each component we explore their stellar mass distributions. \Fig{SSDP} shows the stellar surface density radial profiles for the total stellar mass (black points) and for the disc (blue points) and spheroid (red points) components for the sample of the 8 galaxies shown in \Fig{SMD} and sorted based on their \DTTdt\ values. These galaxies represent a blend of disc-dominated (\DTTdt\ > 0.5) and spheroid-dominated (\DTTdt\ < 0.5) galaxies. We perform a two-component fit of the face-on total stellar surface density by using a non-linear least square method to fit a double \cite{S68} profile (black line) which consists of an one-dimensional Sersic function and a fixed $n=1$ exponential profile
\begin{align} \label{eq:DeltaJgas}
\Sigma (r) = \Sigma_{0,\mathrm{s}}\, \exp \left[-b_n \left(\frac{r}{R_\mathrm{eff}} \right)^{1/n} \right] + \Sigma_{0,\mathrm{d}}\, \exp\left[-\left( \frac{r}{R_\mathrm{d}} \right) \right]\; ,
\end{align}
where $r$ is the projected 2-D radius, $\Sigma_{0,\mathrm{s}}$ and $\Sigma_{0,\mathrm{d}}$ are the central surface densities of the spheroid and disc, respectively, $b_n$ is the Sersic coefficient parameter, $n$ is the Sersic index, $R_\mathrm{eff}$ is the effective radius that encloses half of the projected total stellar mass and $R_\mathrm{d}$ is the disc scale length (see also \App{B} for more details). In addition, we fit independently the disc with an exponential profile (blue curve) and the spheroid with a Sersic profile (red curve).

Even though we do not restrict the physical extent of our components, we see a clear trend which holds for all galaxies in \Fig{SSDP} and shows that the spheroid component dominates the mass budget in the central regions even for the disc-dominated galaxies; while we notice a drop in disc's surface density in the central regions. This is in agreement with previous works \citep[e.g.,][]{ODB13,BPG20b} who argued that disc components exhibit central intensity depression which lowers their contribution to the stellar mass. This drop indicates that it is easier for particles to scatter away from small \Dt\ near the centre (i.e., based on our method they do not have disc-like kinematics). In most cases there is an increase in disc's surface density between 3 and 7 kpc (right after the aforementioned drop) which partially exists due to contamination between the components. 

In galaxy 25 we see that at r$\sim$20kpc there is a prominent bump which appears both in the total (black points) and the disc (blue points) surface densities. This excess of stellar mass is related to spiral arms and reinforces our finding that galaxy 25 has spiral arms which our method correctly associates with its disc component (see \protect\Fig{SSD}). In general, the majority of spheroids tightly follow a Sersic profile especially in the central regions, while discs are well fit by an exponential profile (even when they are sub-dominant components) particularly at their outskirts, as expected.
\section{Results} \label{sec:Results}
Having detailed the methodology and classification of our components, we present in this section relations between \DTTdt\ and galactic/component properties for \eagle\ galaxies selected as described in \Sec{method:The galaxy sample}.

\subsection{Correlations with galactic properties} \label{sec:results:Correlations with galactic properties}
\begin{figure*}
\centering \includegraphics[width=\figdblsize]{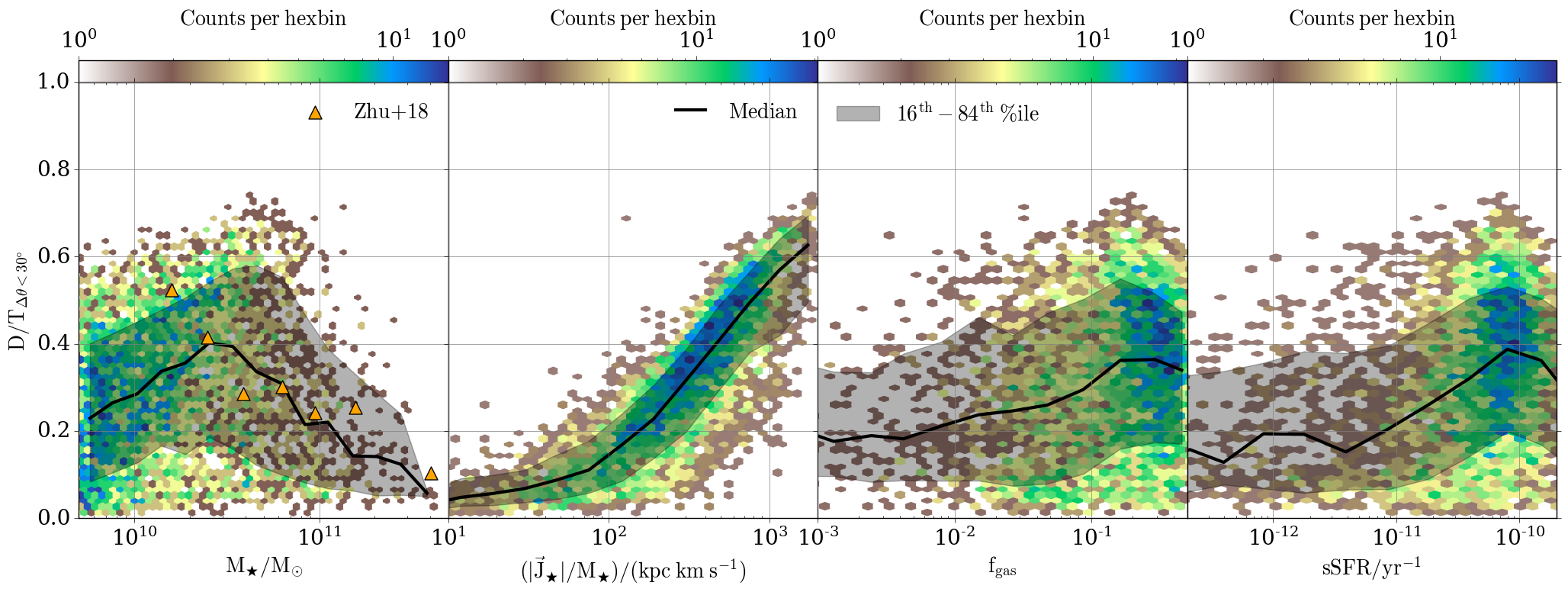}
\caption{The \DTTdt\ ratio as a function of \textbf{first panel:} stellar mass, \textbf{second panel:} stellar specific angular momentum, \textbf{third panel:} gas fraction and \textbf{fourth panel:} specific star formation rate. In each panel the black line and shaded region represent the median and $\mathrm{16^{th}-84^{th}}$ percentile range, respectively. The orange triangles in the first panel represent galaxies from \protect\cite{ZVV18}.} 
\label{fig:DTT_GP}
\end{figure*}

In this work we use the information depicted on angular momentum maps to estimate the relative mass contribution of each component to the total stellar mass. Hence, in this section we use the \DTTdt\ ratio and study its correlations with galactic properties.

\Fig{DTT_GP} shows, from left to right, the dependence of the \DTTdt\ ratio on stellar mass, stellar specific angular momentum, gas fraction (gas to gas+stellar mass) and specific star formation rate (sSFR). In the first panel we compare with the \cite{ZVV18} \citep[300 nearby Calar Alto Legacy Integral Field Area (CALIFA)][galaxies]{SKG12} dataset as presented in \cite{TDH19} (see their Section 3.2). Intermediate mass ($2\times10^{10} \lesssim M_{\star}/M_{\odot}\lesssim 4\times10^{10}$) galaxies have the highest \DTTdt\ ratios while as we move to higher or lower masses, galaxies tend to be more spheroid-dominated - in agreement with the observational data. A similar behaviour has been reported before for the \eagle\ \citep[see e.g., Figure 4 of][]{CSF18} and the IllustrisTNG \citep[see e.g., Figure 3 of][]{TDH19} simulations. In addition, observational studies \citep[e.g.,][]{MLD16,TSB16} have also reported broadly similar relations.

In addition, we see a tight (positive) correlation between the \DTTdt\ ratio and the specific angular momentum of the galaxy, which indicates that the higher the disc contribution to the total stellar mass is the faster the galaxy is rotating \citep{DTK20}. This is an expected behaviour since our method by construction will assign higher \DTTdt\ ratios to galaxies with higher angular momenta.

Lastly, we find a weak trend between both the gas fraction and sSFR with \DTTdt, showing that more gas rich and star forming galaxies have preferentially more prominent disc components. These conclusions are in agreement with \cite{LSB18b} who found that (at fixed stellar mass) passive galaxies have lower spin parameters than star-forming ones.

\subsection{Correlations with other methods} \label{sec:results:Correlations with other methods}
\begin{figure*}
\centering \includegraphics[width=\figdblsize]{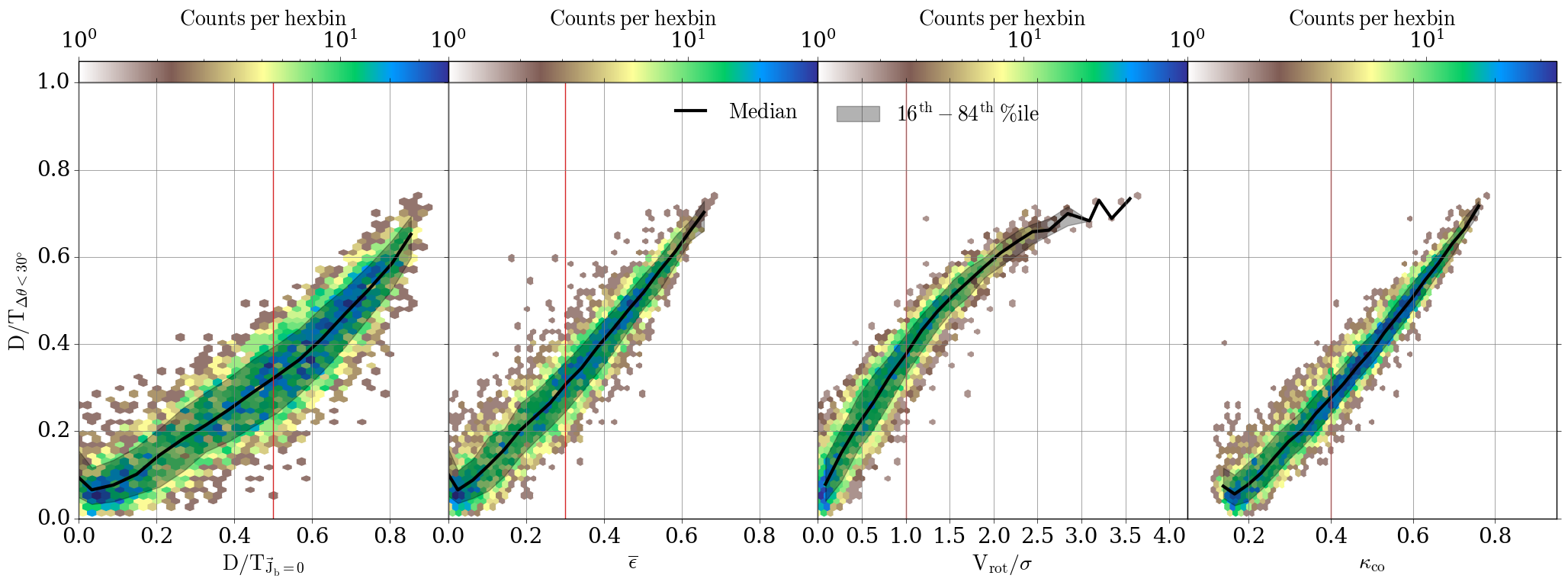}
\caption{The \DTTdt\ ratio as a function of \textbf{first panel:} \DTTj\ ratio, \textbf{second panel:} average circularity parameter ($\overline{\epsilon}$), \textbf{third panel:} ratio between rotation and dispersion velocities ($v_{\mathrm{rot}}/\sigma$) and \textbf{fourth panel:} co-rotation fraction ($\kappa_{\mathrm{co}}$). In each panel the black line and shaded region represent the median and $\mathrm{16^{th}-84^{th}}$ percentile range, respectively. The red vertical lines represent the limits usually used \protect\citep[see e.g.,][]{CSC17,TCM19} to dichotomise galaxies into late-type and early-type (values higher and lower than the red lines, respectively).} 
\label{fig:DTT_MP}
\end{figure*}

Even though it was originally assumed that early-type galaxies (ETGs) and (classical) bulges are solely pressure-supported systems, it was later revealed that they often do have some rotation \cite[e.g.,][]{BC75,KI82,D83,SB95,ECK11}. In general, galactic spheroids and ETGs are dominated by stars in low angular momentum orbits, while late-type galaxies (LTGs) are populated by stars performing ordered rotational orbits. Theoretical studies usually use the D/T ratio to distinguish between ETGs and LTGs, hence the method used to perform the decomposition should be able to accurately capture the distinct kinematic components.

For that reason, we compare in \Fig{DTT_MP} the \DTTdt\ ratio with, from left to right, the \DTTj\ ratio, the average circularity parameter ($\overline{\epsilon}$), the ratio between rotation and dispersion velocities ($v_{\mathrm{rot}}/\sigma$), and the fraction of stellar particles' kinetic energy invested in co-rotation \citep[$\kappa_{\mathrm{co}}$ introduced by][]{CSC17}. We find positive correlations between our method and the aforementioned ones, with the \DTTj\ ratio showing the most scatter and $\kappa_{\mathrm{co}}$ and $v_{\mathrm{rot}}/\sigma$ the least. The latter curves over towards high $v_{\mathrm{rot}}/\sigma$ values in disc-dominated systems, as has been recently reported by \cite{TCM19}.

We note that the \DTTdt\ value where the demarcation (red) lines meet the median (black) lines lies lower than the 0.5 value used by some other methods \citep[e.g.,][]{RPT18,TRB19,RTP19} to separate disc-dominated from bulge-dominated galaxies;
and it is closer to that of, for example, \cite{LCP08,WJK09,GCP15,PT15,ITH19,OEL20,ZSL20}. Finally, we note that even though the four methods form tight relations with \DTTdt, on some occasions we see a mismatch between the predictions which can lead in miss-classifications. These disagreements are responsible for the dispersion seen in \Fig{DTT_MP} and are more prominent in the second panel, where a few ($\sim$20) galaxies with 2<$\overline{\epsilon}$<3 appear well above the median and the $\mathrm{16^{th}-84^{th}}$ percentile range. Such case is galaxy 25 (presented in \Sec{classification:A representative sample}) which has $\overline{\epsilon}\sim$ 0.2 and \DTTdt $\sim$ 0.5: hence the former method classifies it as an ETG and the latter as disc-dominated.

The fact that our method not only identifies such discrepancies but also provides a deeper insight and is able to explain them is what makes it unique. In addition, it differs from some previous methods since it does not contain any a priori assumptions regarding the kinematics of the components ( e.g. that stellar particles counter-rotating with respect to the total angular momentum are bulge particles) hence, provides a more physically-motivated way of identifying kinematically distinct components.

\subsection{Correlations with environment} \label{sec:results:Correlations with environment}
\begin{figure*}
\centering \includegraphics[width=\figdblsize]{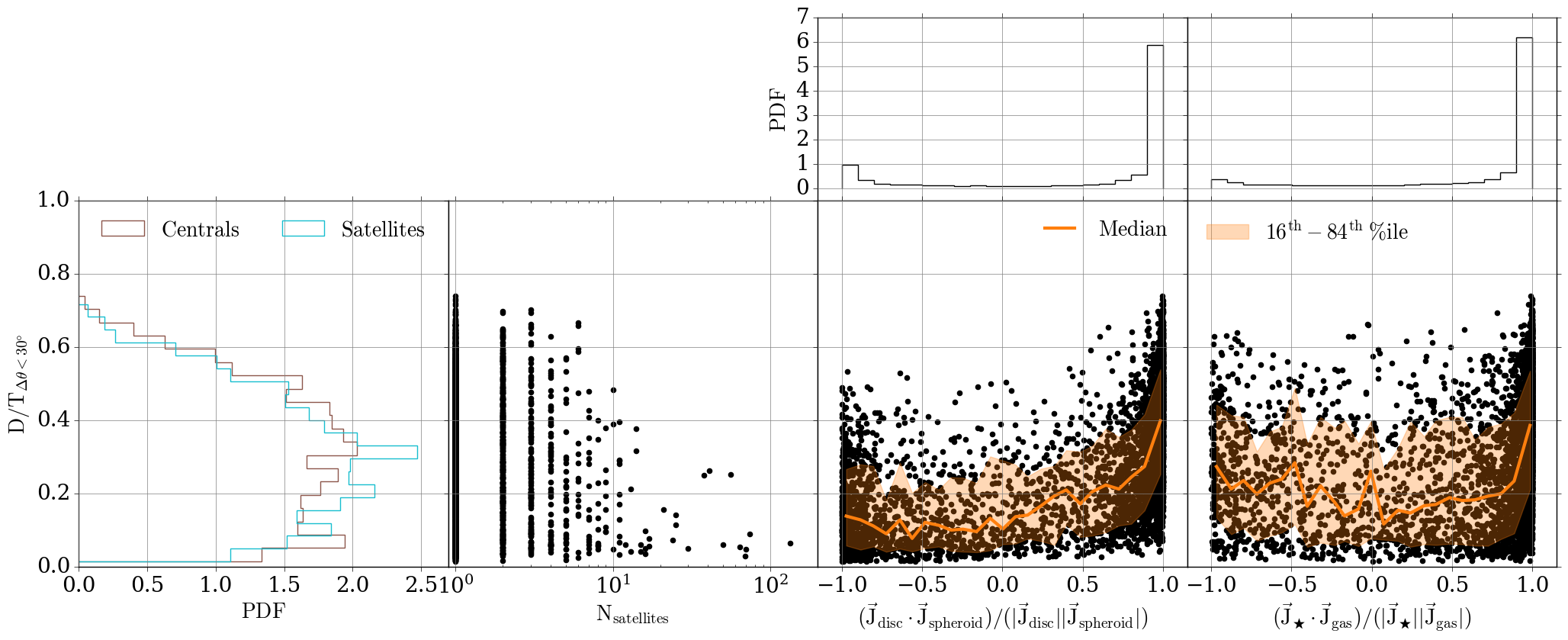}
\caption{\textbf{First panel:} the \DTTdt\ ratio probability density function for central (brown) and satellite (cyan) galaxies and the \DTTdt\ ratio as a function of \textbf{second panel:} number of satellites (i.e., number of galaxies that share the same group number), \textbf{third panel:} misalignment between the angular momentum of the disc and the spheroid components, and \textbf{fourth panel:} misalignment between the angular momentum of the stellar and the gaseous components. In the last two panels we attached two panels on top showing the probability density function of the x-axis data. In addition, the orange lines and shaded regions represent the median and $\mathrm{16^{th}-84^{th}}$ percentile range, respectively.} 
\label{fig:DTT_E}
\end{figure*}

Dense environments are prone to frequent mergers \citep{NOE14,TDH19} which are the natural culprits for converting ordered motion to random orbits and lowering the angular momentum of the remnant \citep{HBC10,LSB18a,JMK20}; this will also affect the \DTTdt\ ratios.

To study this effect, \Fig{DTT_E} shows, from left to right, the probability density function for central and satellite galaxies \citep[i.e., galaxies with subgroup number zero and greater than zero, respectively, based on the \eagle\ nomenclature][]{MHS16} and the dependence of the number of satellites, the misalignment between the angular momentum of the disc and the spheroid components, and of the stellar and the gaseous (defined following \Eq{j_i} and \Eq{j_star} but using gas particles instead of stellar) components on the \DTTdt\ ratio. Central and satellite galaxies share relatively similar distributions of \DTTdt\ ratios with preferred values between 0.2 and 0.5. However, there is a clear trend between the \DTTdt\ ratios and the number of satellites which indicates that the more satellites a central galaxy has the more dominating its spheroid component is (i.e., lower \DTTdt\ values); the well-know morphology-density relation. As far as the misalignment between the disc and spheroid components is concerned, the median line reveals a weak trend (which becomes more significant as the angular momentum vectors align) which indicates that aligned/counter-rotating components exist in more disc/spheroid-dominated galaxies. Similar behaviour is present for the stellar and gaseous components, however with an additional weak trend as the two become more anti-aligned. The alignment (or not) between disc and spheroid, and stellar and gaseous components is linked to environmental effects \citep{SWS09,SNT12,AWN14,GHW18,PYD19}, and we intend to further investigate this behaviour in a future work.

\subsection{Mass-specific angular momentum relation} \label{sec:results:Mass-specific angular momentum relation}
\begin{figure}
\centering \includegraphics[width=\figsize]{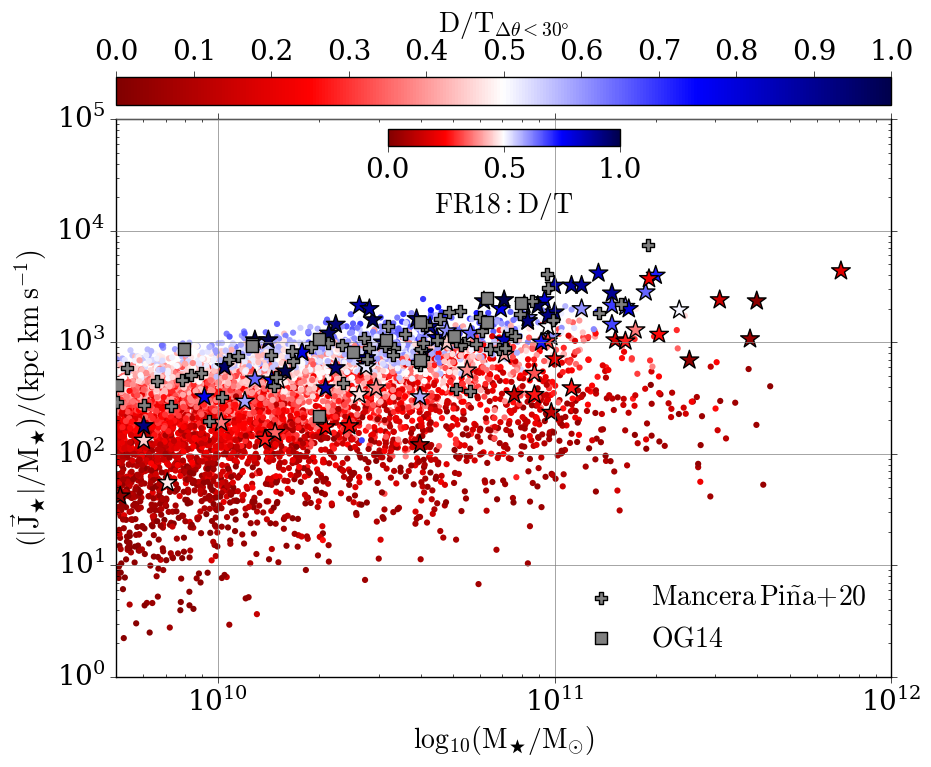}
\caption{The stellar specific angular momentum as a function of the stellar mass colour-coded by the \DTTdt\ ratio. The gray squares and crosses represent galaxies from \protect\cite{OG14} and \protect\cite{MPPF20}, respectively, and the stars represent galaxies from \protect\cite{FR18} colour-coded by their D/T ratio.} 
\label{fig:AM_M}
\end{figure}

\begin{figure}
\centering \includegraphics[width=\figsize]{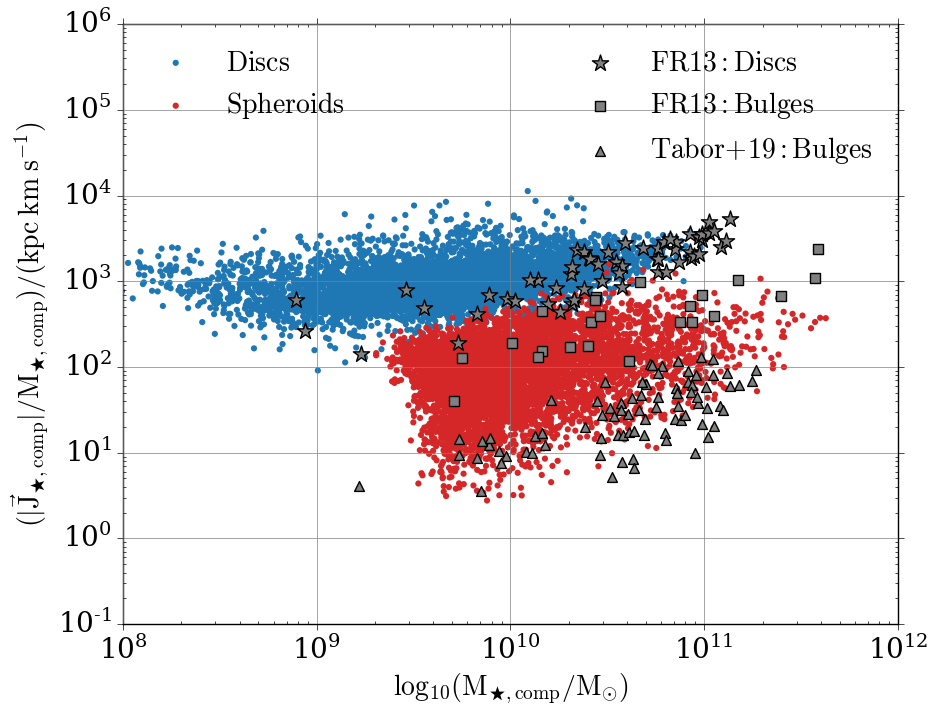}
\caption{The component specific angular momentum as a function of the component stellar mass. The gray stars and squares represent discs and bulges, respectively, from \protect\cite{FR13}, and the gray triangles represent bulges from \protect\cite{TMA19}.} 
\label{fig:C_AM_M}
\end{figure}

The angular momentum of a galaxy determines its size, morphology and contains information regarding its formation path \citep{CFB16,PFD18,SFG18}. The relative contribution of the disc and spheroid components to the total mass has been shown to dictate where a galaxy lies on the mass-specific angular momentum plane \citep{RF12,FR18,TMA19}. 

\Fig{AM_M} shows the stellar specific angular momentum as a function of the stellar mass colour-coded by the \DTTdt\ ratio of each galaxy. We compare our results with the following observational data sets:
\begin{enumerate}
\item \cite{OG14}:\,a sample of 16 nearby spiral galaxies from the The HI Nearby Galaxy Survey (THINGS) \citep{WBD08} selected based on their Hubble types T (from Sab to Scd) which were taken from the HyperLeda database \citep{PPP03}.
\item \cite{FR18}:\,a sample of 57 spirals, 14 lenticulars and 23 ellipticals decomposed following a two-dimensional decomposition of r-band images with a simple exponential and de Vaucouleurs profiles \citep{K86,K87,K88}.
\item \cite{MPPF20}:\,a large sample of nearby disc galaxies which contains 90 galaxies from the Spitzer Photometry \& Accurate Rotation Curves (SPARC) \citep{LMS16}, 30 from the \cite{PVB16} sample, 16 from the Local Irregulars That Trace Luminosity Extremes, The Hi Nearby Galaxy Survey (LITTLE THINGS) \citep{HFA12}, 14 from the Local Volume Hi Survey (LVHIS) \citep{KWK18}, 4 from the Very Large Array survey of Advanced Camera for Surveys Nearby Galaxy Survey Treasury galaxies (VLA-ANGST) \citep{OSW12} and 3 from the Westerbork HI Survey of Irregular and Spiral Galaxies (WHISP) \citep{VVS01}.
\end{enumerate}
We match extremely well both the tight mass-angular momentum relation and the \DTTdt\ dependent vertical colour gradient which implies that for a fixed stellar mass disc-dominated galaxies have higher angular momenta than spheroid-dominated ones.

\Fig{C_AM_M} shows the specific angular momentum separately for the disc (red) and spheroid (blue) components as a function of their stellar mass. We compare our results with the following observational data sets:
\begin{enumerate}
\item \cite{FR13}:\,a sample of nearby bright galaxies of all types introduced in \cite{RF12}.
\item \cite{TMA19}:\, a sample of early-type galaxies from the Mapping Nearby Galaxies at Apache Point Observatory (MaNGA) \citep{BBL15,DMB15} selected from the Galaxy Zoo 2 catalogue \citep{WLB13} as objects with ‘smooth’ vote fraction > 0.7.
\end{enumerate}

We find an adequate agreement both with \cite{FR13} and \cite{TMA19} data sets. Similar to the disc-dominated galaxies in \Fig{AM_M}, the disc components follow a tight relation which is more constrained than the one spheroids follow. The wide scatter in the spheroid specific angular momentum can be also seen in the observational data where the two surveys cover different locations on the plot. This behaviour potentially reflects selection effects which bias the \cite{TMA19} sample in favour of low angular momentum objects since they studied early-type galaxies, whereas \cite{FR13} analysed galaxies of different morphological types \citep{RF12}. Furthermore, disagreements between the two surveys are also expected since they deduced D/T ratios following different techniques. 

\subsection{The baryonic Tully-Fisher and Faber-Jackson relations} \label{sec:results:The baryonic Tully-Fisher and Faber-Jackson relations}
\begin{figure}
\centering \includegraphics[width=\figsize]{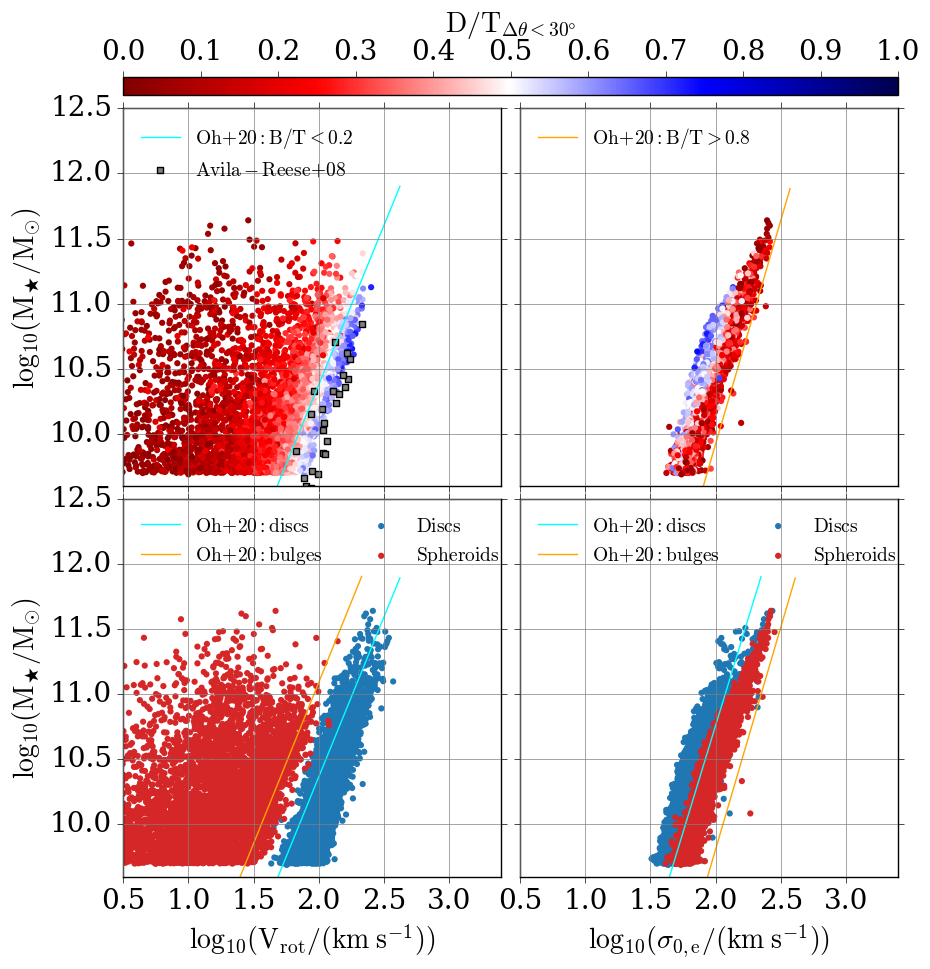}
\caption{The baryonic Tully-Fisher (left column) and Faber-Jackson (right column) relations. \textbf{Top left panel:} The galactic stellar mass as a function of the galactic rotational velocity colour-coded by the \DTTdt\ ratio. The gray squares and cyan line represent disc-dominated galaxies from \protect\cite{AZF08} and \protect\cite{OCB20}, respectively. \textbf{Bottom left panel:} The galactic stellar mass as a function of the disc (blue) and spheroid (red) rotational velocity. Cyan and orange lines represent the disc and bulge components, respectively, from \protect\cite{OCB20}. \textbf{Top right panel:} The galactic stellar mass as a function of the galactic line-of-sight velocity dispersion colour-coded by the \DTTdt\ ratio. The orange line represents bulge-dominated galaxies from \protect\cite{OCB20}. \textbf{Bottom right panel:} The galactic stellar mass as a function of the disc (blue) and spheroid (red) line-of-sight velocity dispersion (calcuated within the half-mass redius). The cyan and orange lines represent the disc and bulge components, respectively, from \protect\cite{OCB20}}
\label{fig:TFFJ}
\end{figure}

The \cite{TF77} and \cite{FJ76} relations reflect fundamental correlations between the stellar mass and kinematics of rotationally-supported LTGs and pressure-supported ETGS, respectively. Given the similarities between the disc and spheroid components with LTGs and ETGs, respectively, we explore if our kinematically distinct components follow the aforementioned relations. 

\Fig{TFFJ} shows the baryonic \citep{MSB00} Tully-Fisher (left column) and Faber-Jackson (right column) relations for each galaxy (top row) and for the disc and spheroid components (bottom row). We compare our results with the following observational data sets:
\begin{enumerate}
\item \cite{AZF08}:\,a sample of non–interacting disc galaxies compiled from the literature \citep{ZAH03}.
\item \cite{OCB20}:\, a sample of 195 ellipticals, 336 lenticulars, and 295 spirals from the Sydney-AAO Multi-object Integral (SAMI) field spectrograph survey \citep{CLB12}.
\end{enumerate}

We adopt the method introduced in \cite{TCM19} to estimate the rotation velocity ($V_{\mathrm{rot}}$) and line-of-sight velocity dispersion ($\sigma_\mathrm{0,e}$) for the whole galaxy and its components, however we only use particles within one half-mass radius when calculating the latter. 

Our galaxies (top left panel) are in great agreement with the observational data of \cite{AZF08} and \cite{OCB20} and show a tight correlation for \DTTdt\ values higher than 0.5 (i.e., for disc-dominated galaxies). This correlation appears to vanish as we move to galaxies with a considerable spheroid component (redder colours), as expected. The degree to which each component affects its host galaxy kinematics becomes clear in the bottom left panel which shows the Tully-Fisher relation for the disc (blue) and spheroid (red) component of each galaxy. The disc components follow a well constrained behaviour, while spheroid components' mass and rotational velocity are almost uncorrelated with a significantly larger scatter than the disc, as also found by \cite{OCB20}. 

Furthermore, the galactic Faber-Jackson relation reveals a tight correlation between the stellar mass and velocity dispersion where galaxies with higher \DTTdt\ have lower $\sigma_\mathrm{0,e}$ due to
their higher degree of ordered rotation. In the bottom right panel we find, in agreement with \cite{DHD02}, that discy (spheroidal) structures are formed from dynamically cold (hot) stellar
particles. However, we notice that, while our results have similar slope to \cite{OCB20}, they show a slight offset in normalisation. A possible explanation can be that \cite{OCB20} calculated the
velocity dispersion as the average flux-weighted velocity dispersion ($\sigma_{0}$) of all spaxels inside the effective radius, where $\sigma_{0}$ of each spaxel was extracted after fitting a Gaussian
line-of-sight velocity distribution; whereas we, following \cite{TCM19}, estimate $\sigma_\mathrm{0,e}$ as the remaining motion (after subtraction the ordered co-rotation component) in the disc plane.

\subsection{Anisotropy parameter} \label{sec:results:Anisotropy parameter}
\begin{figure}
\centering \includegraphics[width=\figsize]{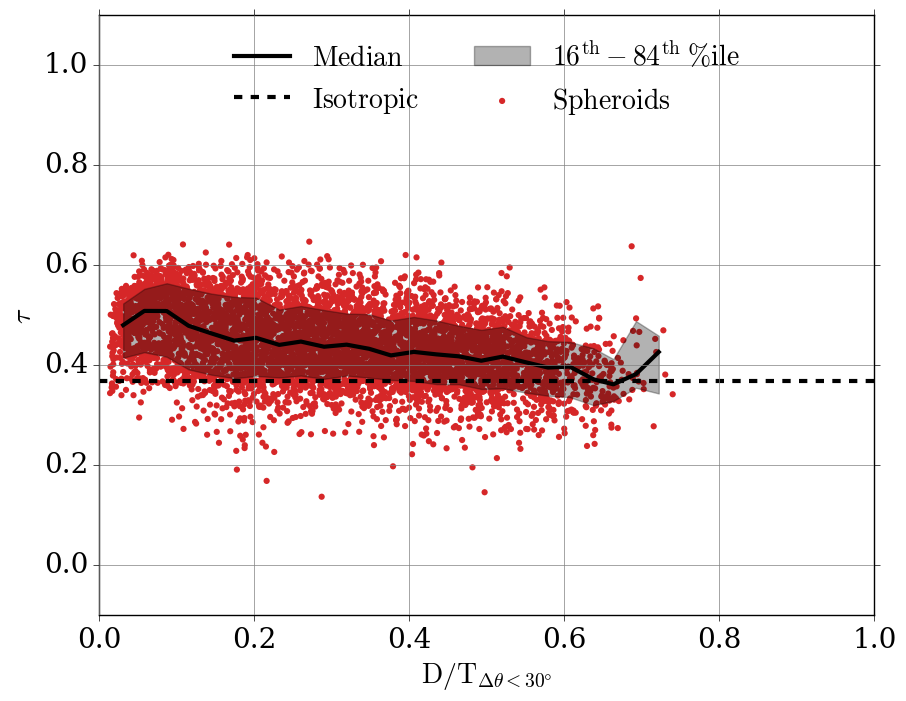}
\caption{Anisotropy parameter for the spheroid components as a function of the \DTTdt\ ratio. The black line and shaded region represent the median and $\mathrm{16^{th}-84^{th}}$ percentile range,
  respectively. The black dashed horizontal line represents isotropic orbits with $\tau=e^{-1}=0.367$.}
\label{fig:C_B_DTT}
\end{figure}

The anisotropy of stellar orbits within a system \citep{BT08} can be quantified by the parameter:
\begin{align} \label{eq:DeltaJgas}
\beta  = 1 - \frac{\overline{u_{\theta}^{2}} + \overline{u_{\phi}^{2}}}{2\overline{u_{r}^{2}}}\; ,
\end{align}
where $\overline{u_{\theta}}, \overline{u_{\phi}}$ and $\overline{u_{r}}$ are the average azimuthal, polar and radial components of the velocity of each stellar particle. Hence, different types of orbits result in different values of $\beta$:
 \[
    \beta = 
		\begin{cases}
        -\inf, & \text{orbits are circular.}\\
        <0, & \text{orbits are tangentially biased.}\\
        =0, & \text{orbits are isotropic.}\\
        >0, & \text{orbits are radially biased.}\\
        1, & \text{orbits are radial.}
        \end{cases}
  \]

In order to study the orbital composition of our spheroids we use a more convenient form of the anisotropy parameter: $\tau \equiv e^{\beta -1}$. Hence, the five cases described above now change to $\tau$=0, 0<$\tau$<0.367, $\tau$=0.367, 0.367<$\tau$<1  and $\tau$=1, respectively. 

\Fig{C_B_DTT} shows $\tau$ as a function of the \DTTdt\ ratio. Stellar particles which form spheroids have slightly tangentially biased orbits for low \DTTdt\ values and gradually move to more isotropic orbits (i.e., $\tau$ values closer to the black dashed line) as \DTTdt\ increases. In other words spheroid-dominated galaxies consist of stellar particles in tangentially biased orbits while spheroids within disc-dominated galaxies consist of stellar particles in isotropic orbits. This trend while present is quite weak hence agrees with the kinematic similarities between spheroid components and ETGs found by previous studies \citep[see][for a review]{K16}.

\subsection{Black-hole spheroid mass relation} \label{sec:results:Black-hole spheroid mass relation}
\begin{figure}
\centering \includegraphics[width=\figsize]{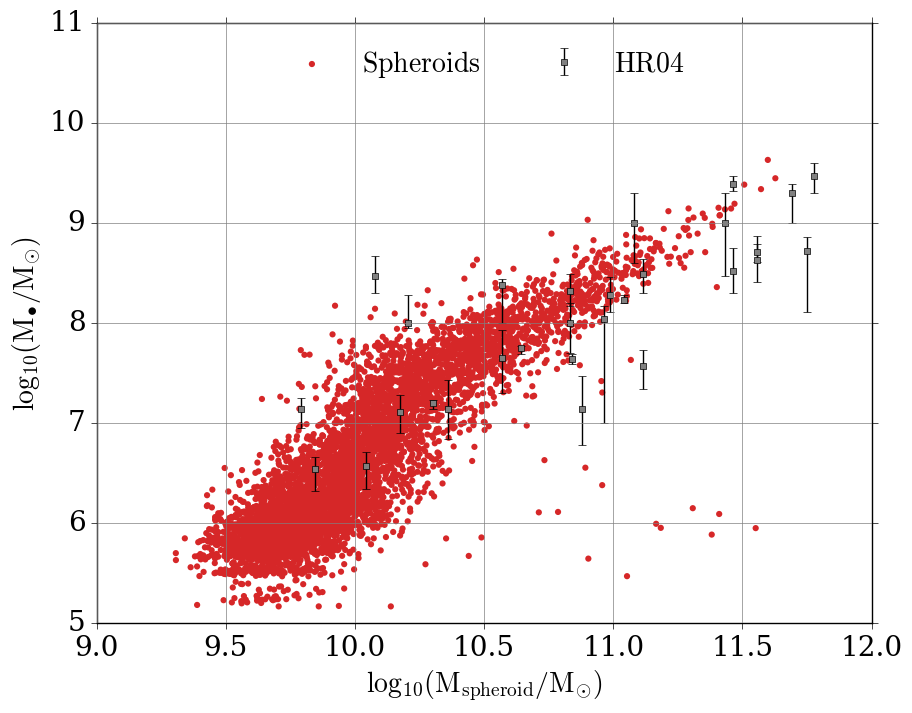}
\caption{The black hole mass as a function of spheroid mass. The gray squares represent \protect\cite{HR04} data.} 
\label{fig:B_B_M}
\end{figure}

Tight relations between black hole and spheroid properties have been revealed in numerous studies and have been linked with the fact that they form in closely associated processes \citep{KG01,MH03,GHB08,GK09,SMH11,SMM12}. 

\Fig{B_B_M} shows the relation between the spheroid and black hole masses\footnote{As discussed in \cite{MHS16,TET17} this is the summed mass of all black holes associated with each subhalo.}. We compare with the \cite{HR04} (sample of 30 nearby galaxies) dataset. We find that our spheroids form an almost linear relation (in log-space) between the black hole and spheroid mass which extends to all masses; a behaviour which is in agreement with previous work \citep{BCC12,G12,MM13,BSB16,HLS20}. Hence, we broadly match the \cite{HR04} relation despite the different definitions and extraction techniques used to estimate the mass of the bulge/spheroid component.

It is worth noting that the free parameters of the \eagle\ black hole model have been calibrated to match the black-hole-bulge mass relation; however that was performed using both different datasets (i.e., \cite{MM13} instead of \cite{HR04}) and different masses (i.e., \cite{SCB15,CSB15} used the total stellar mass instead of the spheroid mass).
  
\subsection{Age and metallicity relations} \label{sec:results:Age and metallicity relations}
\begin{figure}
\centering \includegraphics[width=\figsize]{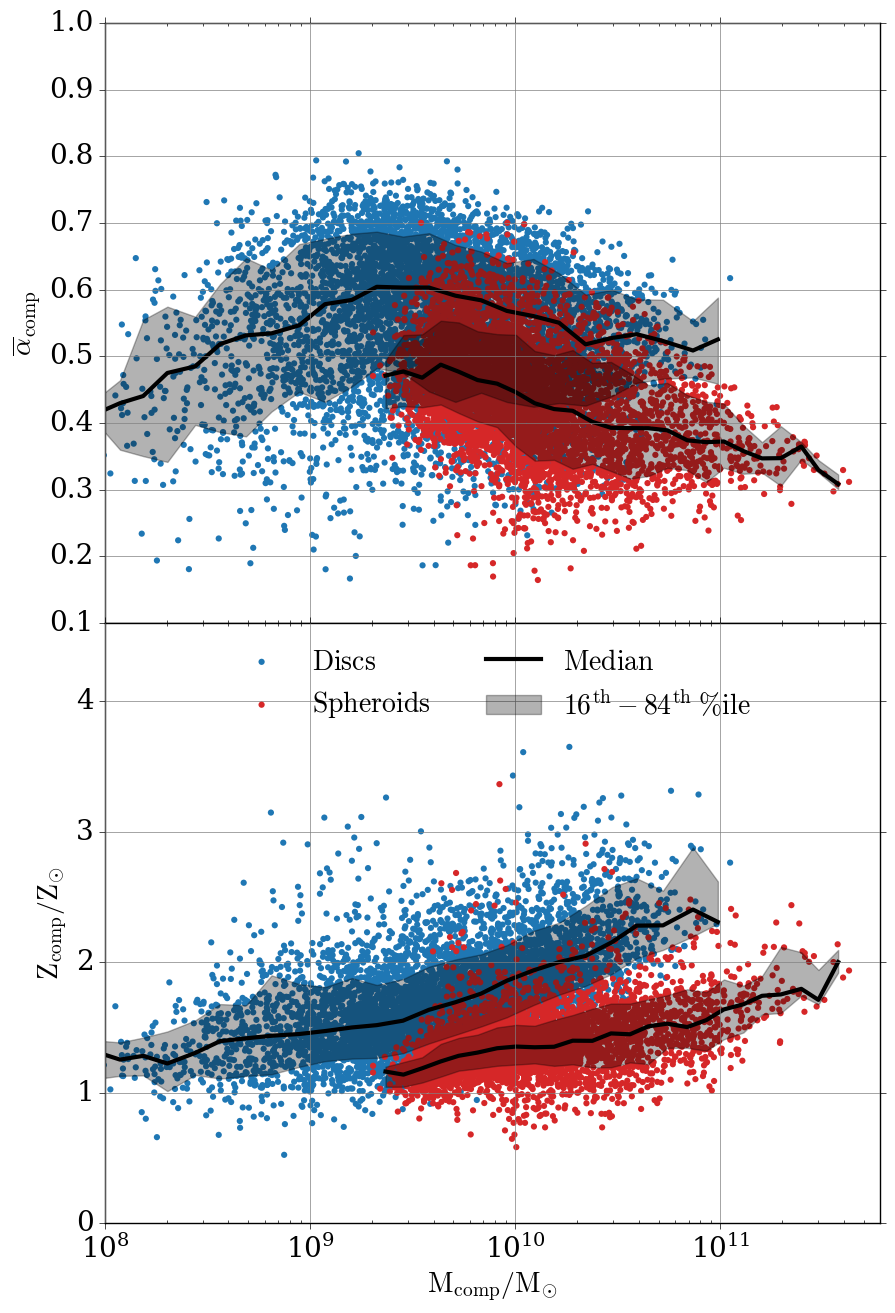}
\caption{The average stellar formation expansion factor (top) and metallicity (bottom) of the disc (blue) and spheroid (red) component as a function of the component stellar mass. The black line and shaded region represent the median and $\mathrm{16^{th}-84^{th}}$ percentile range, respectively.} 
\label{fig:C_A_M_M}
\end{figure}

The age and metallicity of galactic components reflect their formation processes and time-scales \citep{G09,ODB13,TMM20}, hence distinct components should follow different relations. 

\Fig{C_A_M_M} shows the mass-age and mass-metallicity relations for the disc and spheroid components. We use $\overline{\alpha}_\mathrm{comp}$ as a proxy for the age of each component, which we define as the average birth (i.e., at the time a stellar particle is born) scale factor for all stellar particles belonging to that component \citep{TET17}. In addition, the metallicity of each component is defined as the sum of the mass-weighted metallicity $Z$ of all stellar particles belonging to that component, where $Z$ is the mass fraction of elements heavier than helium. At a given component stellar mass, there is a clear trend (based on the median lines) which shows that disc components are younger and more metal-rich than the spheroid components, in agreement with previous simulations \citep[e.g.,][]{NOE14,PYD19,RTP19,WOL19} and observations \citep[e.g.,][]{BG90,MDJ19}.

\section{Discussion} \label{sec:Discussion}
\begin{figure}
\centering \includegraphics[width=\figsize]{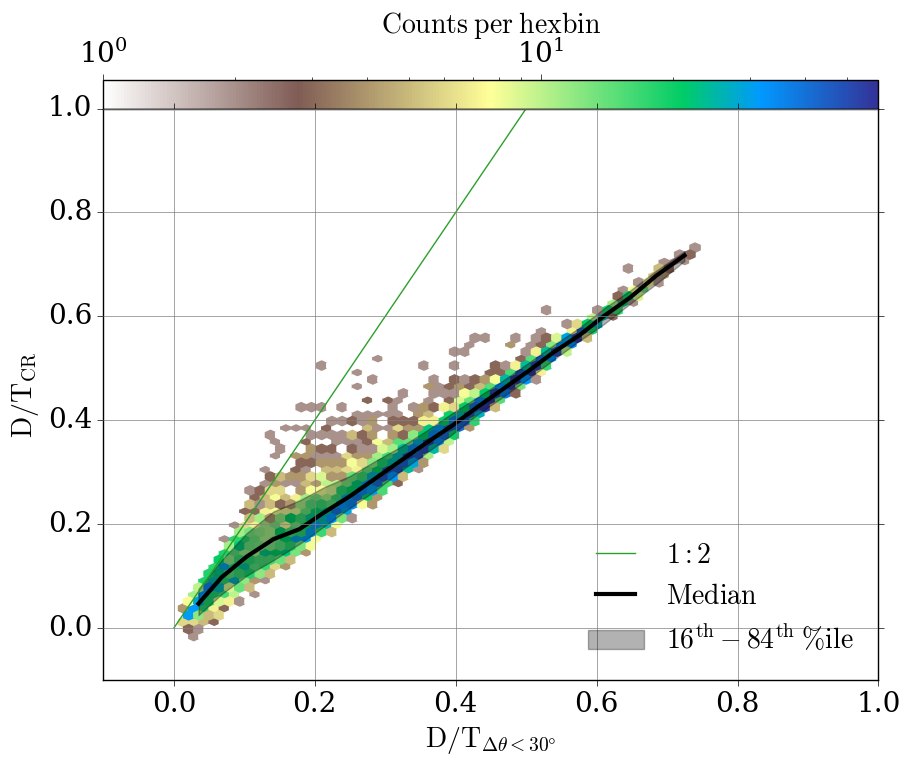}
\caption{The \DTTcr\ ratio as a function of the \DTTdt\ ratio. The black line and shaded region represent the median and $\mathrm{16^{th}-84^{th}}$ percentile range, respectively and the green line represents the 1:2 ratio.}
\label{fig:DTTCR_DTT}
\end{figure}

\begin{figure*}
\centering \includegraphics[width=\figdblsize]{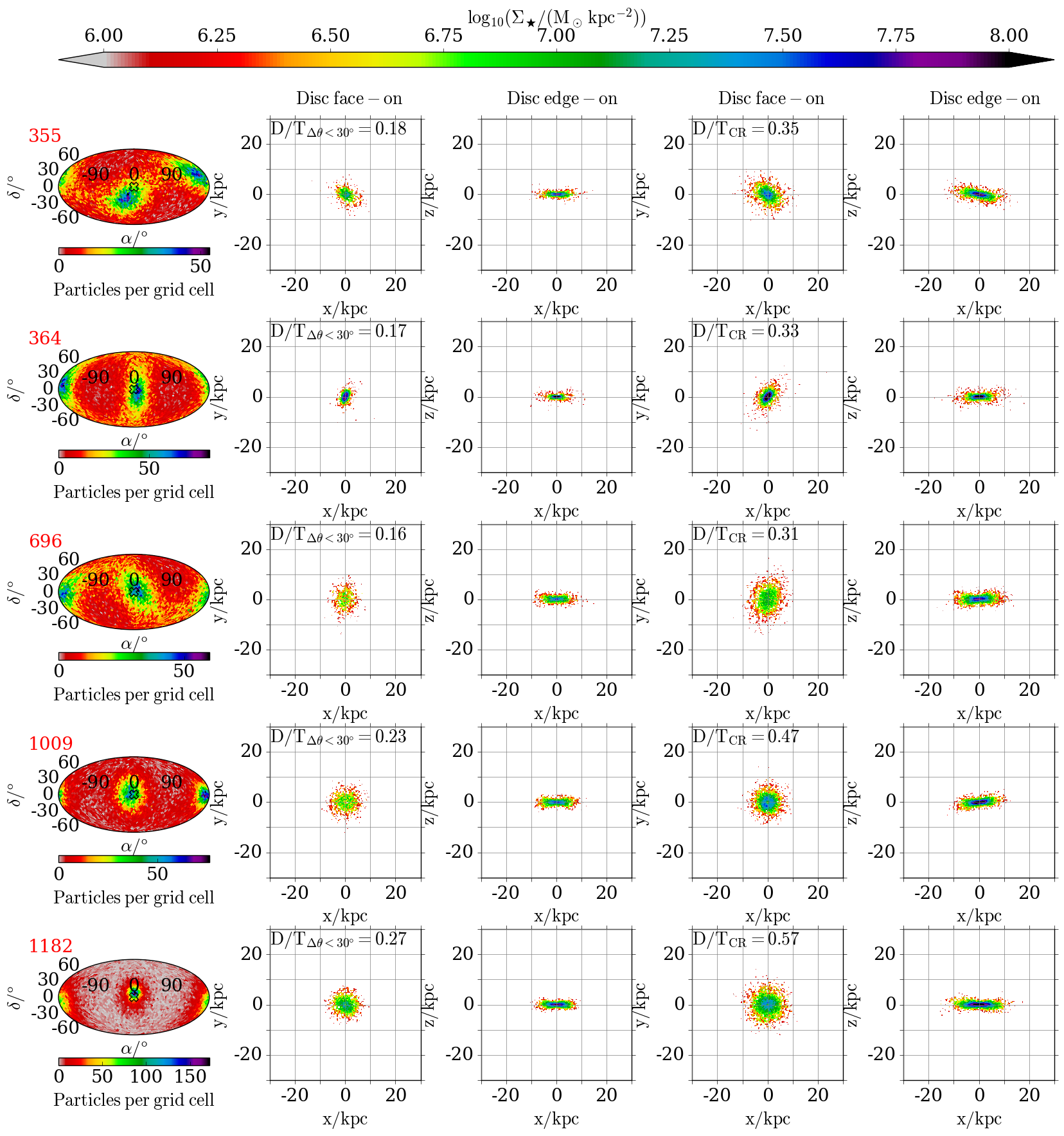}
\caption{A sample of 5 galaxies selected so that each one has a ratio of \DTTcr\ to \DTTdt\ higher than 2. Each galaxy is represented by five panels as described below. The first column displays for each galaxy on a Mollweide projection the angular momentum maps, where the red number on the top left corner in each row refers to each galaxy's group number. As in \Fig{SRAEl} the galaxies have been rotated so that \Jstar\ (hollow X symbol) points towards the reader. The second and third columns have the face-on and edge-on projections, respectively, for the disc component when it is defined based the method described in \Sec{method:Decomposition}. The fourth and fifth columns have the same projections, respectively, but when the disc component also includes counter-rotating particles (see text for more details). The text in the each panel on the second and fourth column represents the \DTTdt\ and \DTTcr\ ratios, respectively, for each galaxy.}
\label{fig:SDSD}
\end{figure*}

In this work we follow the method introduced in \Sec{method:Decomposition} which decomposes galaxies by identifying kinematically distinct components. This method in order to be in agreement with previous observational and/or theoretical methods assigns to the disc component particles that have angular momenta broadly aligned with the galactic one (the angle between the two should not exceed 30$\degree$, see \Eq{criterion}). However, as discussed in \Sec{classification:A representative sample} and shown in \Fig{DTT_E} there are numerous galaxies whose anti-aligned particles also form a component which counter-rotates (with respect to the galactic angular momentum). It is clear that if these particles are also assigned to the disc component that will have a significant impact on the predicted D/T ratio. In this section we briefly investigate what effect such a method will have on our \DTTdt\ ratios. 

We follow exactly the same process as the one described in \Sec{method:Decomposition} however in step (v) we assign to the disc component particles with
\begin{align} \label{eq:criterion2}
\Dt_{i} < 30 \degree
\end{align}
and
\begin{align} \label{eq:criterion3}
\Dt_{i} > 150 \degree \;.
\end{align}
This results in a new D/T ratio which we call \DTTcr. In addition, we amend step (vi) and we apply to \DTTcr\ a correction with two times $\chi$, to account for the second 30$\degree$ patch on the angular momentum sphere.

\Fig{DTTCR_DTT} shows \DTTcr\ as a function of \DTTdt. The majority of galaxies, especially the ones with \DTTdt\ values higher than 0.2, follow tightly the 1:1 relation as indicated by the median line and 1-$\sigma$ region. Hence, in cases where there is no significant counter-rotating component the \DTTdt\ values are not significantly altered\footnote{Small fluctuations of order $10^{-2}$ between the two ratios are also expected due to the smoothing process which finds and uses grid cell centres within 30$\degree$. So there are particles within 30$\degree$ which are not included in the smoothing but are used when estimating the mass of the component.}. However, some galaxies, especially the low \DTTdt\ ones, have a dramatic change in their D/T ratios that is reflected on their \DTTcr\ values which can be as high as $\sim$ 2.8 times their original \DTTdt\ ratios.

\Fig{SDSD} shows the angular momentum maps and the face-on and edge-on projections for the two different definitions of the disc component for a sample of 5 galaxies who display noticeable differences between their \DTTdt\ and \DTTcr\ ratios. It becomes apparent from the angular momentum maps of all 5 galaxies that two structures exist; one well-aligned with the galactic angular momentum and the other counter-rotating with respect to the latter. From the second and third columns we see that these galaxies have thin rotationally-supported discs which they continue to exist (although increase their height and diameter) when particles with \Dt >150$\degree$ are included. The most interesting case is galaxy 1182 (fifth row) which has \DTTdt\ value of 0.27 while its \DTTcr\ ratio is 0.57. Hence, in with the former definition is classified as a spheroid-dominated galaxy while with the latter a disc-dominated.
\section{Conclusions} \label{sec:Conclusions}
Being able to extract galactic components is vital in studying the galaxy as a whole. In this work we use the information depicted on angular momentum maps (\Sec{Methodology}) to identify kinematically distinct components and study the imprint each component leaves on the properties of their host galaxy. We consider this method a useful addition to the literature since it results in components which are in great agreement with the observed. Our main conclusions are as follows:
\begin{itemize}
\item We find a clear separation in angular momentum space of distinct components and classify galaxies into several morphological types (\Sec{Classification}).
\item We demonstrate that even though we make no assumptions regarding the spatial extend of our components, they follow the expected spatial distribution and surface density profiles (\Sec{classification:A representative sample}).
\item We compare with other methods and find, in general, tight relations but also some interesting cases in which our method provides better classification: for example disc-like galaxies with particles on non-circular orbits. (\Sec{results:Correlations with other methods}).
\item Our method identifies a significant population of galaxies with counter-rotating discs embedded within them (\Sec{results:Correlations with environment}). 
\item Our components have distinct angular momenta (\Sec{results:Mass-specific angular momentum relation}), kinematics (\Sec{results:The baryonic Tully-Fisher and Faber-Jackson relations} and \Sec{results:Anisotropy parameter}) and ages and chemical compositions (\Sec{results:Age and metallicity relations}). Thus our morphological classification reproduces the observed properties of such systems.
\end{itemize}

In a future work we will use our method to study merging systems and galaxies with counter-rotating discs in order to better understand their formation path and evolution. In addition, we intend to explore possible imprint of other kinematically distinct components, such as bars and pseudo-bulges, on the angular momentum maps. 

\section*{Acknowledgements}
The authors contributed in the following way to this paper. DI undertook the vast majority of the coding and data analysis, and produced the figures. PAT supervised DI and together they drafted the paper. DI would like to acknowledge his family members for their financial support.

This work used the DiRAC@Durham facility managed by the Institute for Computational Cosmology on behalf of the STFC DiRAC HPC Facility (www.dirac.ac.uk). The equipment was funded by BEIS capital funding via STFC capital grants ST/K00042X/1, ST/P002293/1, ST/R002371/1 and ST/S002502/1, Durham University and STFC operations grant ST/R000832/1. DiRAC is part of the National e-Infrastructure. 

We thank the developers of Astropy \citep{A13,A18}, HEALPix/healpy \citep{GWH99,GBH02,ZSL19}, Matplotlib \citep{H07}, NetworkX \citep{HSS08}, Numpy \citep{VCG11}, Pandas \citep{M10} and SciPy \citep{VGO20}.

Part of this work was carried out during the COVID-19 pandemic and would not have been possible without the tireless efforts of the essential workers, who did not have the safety of working from their homes.

\section*{Data Availability}
A Python version of the method introduced in this article can be found at \href{https://github.com/DimitriosIrodotou/Irodotou-Thomas-2021}{https://github.com/DimitriosIrodotou/Irodotou-Thomas-2021}. The EAGLE data is publicly available and accessible at \href{http://icc.dur.ac.uk/Eagle/database.php}{http://icc.dur.ac.uk/Eagle/database.php}. The data underlying this article will be shared on reasonable request to the corresponding author.

\bibliographystyle{mnras}
\bibliography{mn-jour,di}

\begin{thebibliography}{}
\makeatletter
\relax
\def\mn@urlcharsother{\let\do\@makeother \do\$\do\&\do\#\do\^\do\_\do\%\do\~}
\def\mn@doi{\begingroup\mn@urlcharsother \@ifnextchar [ {\mn@doi@}
  {\mn@doi@[]}}
\def\mn@doi@[#1]#2{\def\@tempa{#1}\ifx\@tempa\@empty \href
  {http://dx.doi.org/#2} {doi:#2}\else \href {http://dx.doi.org/#2} {#1}\fi
  \endgroup}
\def\mn@eprint#1#2{\mn@eprint@#1:#2::\@nil}
\def\mn@eprint@arXiv#1{\href {http://arxiv.org/abs/#1} {{\tt arXiv:#1}}}
\def\mn@eprint@dblp#1{\href {http://dblp.uni-trier.de/rec/bibtex/#1.xml}
  {dblp:#1}}
\def\mn@eprint@#1:#2:#3:#4\@nil{\def\@tempa {#1}\def\@tempb {#2}\def\@tempc
  {#3}\ifx \@tempc \@empty \let \@tempc \@tempb \let \@tempb \@tempa \fi \ifx
  \@tempb \@empty \def\@tempb {arXiv}\fi \@ifundefined
  {mn@eprint@\@tempb}{\@tempb:\@tempc}{\expandafter \expandafter \csname
  mn@eprint@\@tempb\endcsname \expandafter{\@tempc}}}

\bibitem[\protect\citeauthoryear{{Abadi}, {Navarro}, {Steinmetz}  \&
  {Eke}}{{Abadi} et~al.}{2003}]{ANS03}
{Abadi} M.~G.,  {Navarro} J.~F.,  {Steinmetz} M.,   {Eke} V.~R.,  2003, \mn@doi
  [\apj] {10.1086/375512}, \href
  {https://ui.adsabs.harvard.edu/abs/2003ApJ...591..499A} {591, 499}

\bibitem[\protect\citeauthoryear{{Algorry}, {Navarro}, {Abadi}, {Sales},
  {Steinmetz}  \& {Piontek}}{{Algorry} et~al.}{2014}]{ANA14}
{Algorry} D.~G.,  {Navarro} J.~F.,  {Abadi} M.~G.,  {Sales} L.~V.,  {Steinmetz}
  M.,   {Piontek} F.,  2014, \mn@doi [\mnras] {10.1093/mnras/stt2154}, \href
  {https://ui.adsabs.harvard.edu/abs/2014MNRAS.437.3596A} {437, 3596}

\bibitem[\protect\citeauthoryear{{Astropy Collaboration} et~al.,}{{Astropy
  Collaboration} et~al.}{2013}]{A13}
{Astropy Collaboration} et~al., 2013, \mn@doi [\aap]
  {10.1051/0004-6361/201322068}, \href
  {https://ui.adsabs.harvard.edu/abs/2013A&A...558A..33A} {558, A33}

\bibitem[\protect\citeauthoryear{{Astropy Collaboration} et~al.,}{{Astropy
  Collaboration} et~al.}{2018}]{A18}
{Astropy Collaboration} et~al., 2018, \mn@doi [\aj] {10.3847/1538-3881/aabc4f},
  \href {https://ui.adsabs.harvard.edu/abs/2018AJ....156..123A} {156, 123}

\bibitem[\protect\citeauthoryear{{Aumer}, {White}  \& {Naab}}{{Aumer}
  et~al.}{2014}]{AWN14}
{Aumer} M.,  {White} S. D.~M.,   {Naab} T.,  2014, \mn@doi [\mnras]
  {10.1093/mnras/stu818}, \href
  {https://ui.adsabs.harvard.edu/abs/2014MNRAS.441.3679A} {441, 3679}

\bibitem[\protect\citeauthoryear{{Avila-Reese}, {Zavala}, {Firmani}  \&
  {Hern{\'a}ndez-Toledo}}{{Avila-Reese} et~al.}{2008}]{AZF08}
{Avila-Reese} V.,  {Zavala} J.,  {Firmani} C.,   {Hern{\'a}ndez-Toledo} H.~M.,
  2008, \mn@doi [\aj] {10.1088/0004-6256/136/3/1340}, \href
  {http://adsabs.harvard.edu/abs/2008AJ....136.1340A} {136, 1340}

\bibitem[\protect\citeauthoryear{{Barber}, {Schaye}, {Bower}, {Crain},
  {Schaller}  \& {Theuns}}{{Barber} et~al.}{2016}]{BSB16}
{Barber} C.,  {Schaye} J.,  {Bower} R.~G.,  {Crain} R.~A.,  {Schaller} M.,
  {Theuns} T.,  2016, \mn@doi [\mnras] {10.1093/mnras/stw1018}, \href
  {https://ui.adsabs.harvard.edu/abs/2016MNRAS.460.1147B} {460, 1147}

\bibitem[\protect\citeauthoryear{{Beifiori}, {Courteau}, {Corsini}  \&
  {Zhu}}{{Beifiori} et~al.}{2012}]{BCC12}
{Beifiori} A.,  {Courteau} S.,  {Corsini} E.~M.,   {Zhu} Y.,  2012, \mn@doi
  [\mnras] {10.1111/j.1365-2966.2011.19903.x}, \href
  {https://ui.adsabs.harvard.edu/abs/2012MNRAS.419.2497B} {419, 2497}

\bibitem[\protect\citeauthoryear{{Bender}}{{Bender}}{1988}]{B88}
{Bender} R.,  1988, \aap, \href
  {https://ui.adsabs.harvard.edu/abs/1988A&A...202L...5B} {202, L5}

\bibitem[\protect\citeauthoryear{{Bertola} \& {Capaccioli}}{{Bertola} \&
  {Capaccioli}}{1975}]{BC75}
{Bertola} F.,  {Capaccioli} M.,  1975, \mn@doi [\apj] {10.1086/153808}, \href
  {https://ui.adsabs.harvard.edu/abs/1975ApJ...200..439B} {200, 439}

\bibitem[\protect\citeauthoryear{{Binney} \& {Tremaine}}{{Binney} \&
  {Tremaine}}{2008}]{BT08}
{Binney} J.,  {Tremaine} S.,  2008, {Galactic Dynamics: Second Edition}

\bibitem[\protect\citeauthoryear{{Bothun} \& {Gregg}}{{Bothun} \&
  {Gregg}}{1990}]{BG90}
{Bothun} G.~D.,  {Gregg} M.~D.,  1990, \mn@doi [\apj] {10.1086/168361}, \href
  {https://ui.adsabs.harvard.edu/abs/1990ApJ...350...73B} {350, 73}

\bibitem[\protect\citeauthoryear{{Breda}, {Papaderos}  \& {Gomes}}{{Breda}
  et~al.}{2020}]{BPG20b}
{Breda} I.,  {Papaderos} P.,   {Gomes} J.-M.,  2020, \mn@doi [\aap]
  {10.1051/0004-6361/202037889}, \href
  {https://ui.adsabs.harvard.edu/abs/2020A&A...640A..20B} {640, A20}

\bibitem[\protect\citeauthoryear{{Bundy} et~al.,}{{Bundy} et~al.}{2015}]{BBL15}
{Bundy} K.,  et~al., 2015, \mn@doi [\apj] {10.1088/0004-637X/798/1/7}, \href
  {https://ui.adsabs.harvard.edu/abs/2015ApJ...798....7B} {798, 7}

\bibitem[\protect\citeauthoryear{{Ciotti}}{{Ciotti}}{1991}]{C91}
{Ciotti} L.,  1991, \aap, \href
  {https://ui.adsabs.harvard.edu/abs/1991A&A...249...99C} {249, 99}

\bibitem[\protect\citeauthoryear{{Ciotti} \& {Bertin}}{{Ciotti} \&
  {Bertin}}{1999}]{CB99}
{Ciotti} L.,  {Bertin} G.,  1999, \aap, \href
  {https://ui.adsabs.harvard.edu/abs/1999A&A...352..447C} {352, 447}

\bibitem[\protect\citeauthoryear{{Clauwens}, {Schaye}, {Franx}  \&
  {Bower}}{{Clauwens} et~al.}{2018}]{CSF18}
{Clauwens} B.,  {Schaye} J.,  {Franx} M.,   {Bower} R.~G.,  2018, \mn@doi
  [\mnras] {10.1093/mnras/sty1229}, \href
  {https://ui.adsabs.harvard.edu/abs/2018MNRAS.478.3994C} {478, 3994}

\bibitem[\protect\citeauthoryear{{Coccato}, {Morelli}, {Pizzella}, {Corsini},
  {Buson}  \& {Dalla Bont{\`a}}}{{Coccato} et~al.}{2013}]{CMP13}
{Coccato} L.,  {Morelli} L.,  {Pizzella} A.,  {Corsini} E.~M.,  {Buson} L.~M.,
   {Dalla Bont{\`a}} E.,  2013, \mn@doi [\aap] {10.1051/0004-6361/201220460},
  \href {https://ui.adsabs.harvard.edu/abs/2013A&A...549A...3C} {549, A3}

\bibitem[\protect\citeauthoryear{{Cook}, {Cortese}, {Catinella}  \&
  {Robotham}}{{Cook} et~al.}{2020}]{CCC20}
{Cook} R. H.~W.,  {Cortese} L.,  {Catinella} B.,   {Robotham} A. S.~G.,  2020,
  arXiv e-prints, \href {https://ui.adsabs.harvard.edu/abs/2020arXiv200302464C}
  {p. arXiv:2003.02464}

\bibitem[\protect\citeauthoryear{{Cooper}, {Parry}, {Lowing}, {Cole}  \&
  {Frenk}}{{Cooper} et~al.}{2015}]{CPL15}
{Cooper} A.~P.,  {Parry} O.~H.,  {Lowing} B.,  {Cole} S.,   {Frenk} C.,  2015,
  \mn@doi [\mnras] {10.1093/mnras/stv2057}, \href
  {https://ui.adsabs.harvard.edu/abs/2015MNRAS.454.3185C} {454, 3185}

\bibitem[\protect\citeauthoryear{{Correa}, {Schaye}, {Clauwens}, {Bower},
  {Crain}, {Schaller}, {Theuns}  \& {Thob}}{{Correa} et~al.}{2017}]{CSC17}
{Correa} C.~A.,  {Schaye} J.,  {Clauwens} B.,  {Bower} R.~G.,  {Crain} R.~A.,
  {Schaller} M.,  {Theuns} T.,   {Thob} A. C.~R.,  2017, \mn@doi [\mnras]
  {10.1093/mnrasl/slx133}, \href
  {https://ui.adsabs.harvard.edu/abs/2017MNRAS.472L..45C} {472, L45}

\bibitem[\protect\citeauthoryear{{Cortese} et~al.,}{{Cortese}
  et~al.}{2016}]{CFB16}
{Cortese} L.,  et~al., 2016, \mn@doi [\mnras] {10.1093/mnras/stw1891}, \href
  {https://ui.adsabs.harvard.edu/abs/2016MNRAS.463..170C} {463, 170}

\bibitem[\protect\citeauthoryear{{Crain}, {McCarthy}, {Frenk}, {Theuns}  \&
  {Schaye}}{{Crain} et~al.}{2010}]{CMF10}
{Crain} R.~A.,  {McCarthy} I.~G.,  {Frenk} C.~S.,  {Theuns} T.,   {Schaye} J.,
  2010, \mn@doi [\mnras] {10.1111/j.1365-2966.2010.16985.x}, \href
  {https://ui.adsabs.harvard.edu/abs/2010MNRAS.407.1403C} {407, 1403}

\bibitem[\protect\citeauthoryear{{Crain} et~al.,}{{Crain} et~al.}{2015}]{CSB15}
{Crain} R.~A.,  et~al., 2015, \mn@doi [\mnras] {10.1093/mnras/stv725}, \href
  {https://ui.adsabs.harvard.edu/abs/2015MNRAS.450.1937C} {450, 1937}

\bibitem[\protect\citeauthoryear{{Croom} et~al.,}{{Croom} et~al.}{2012}]{CLB12}
{Croom} S.~M.,  et~al., 2012, \mn@doi [\mnras]
  {10.1111/j.1365-2966.2011.20365.x}, \href
  {https://ui.adsabs.harvard.edu/abs/2012MNRAS.421..872C} {421, 872}

\bibitem[\protect\citeauthoryear{{Davies}, {Efstathiou}, {Fall}, {Illingworth}
  \& {Schechter}}{{Davies} et~al.}{1983}]{D83}
{Davies} R.~L.,  {Efstathiou} G.,  {Fall} S.~M.,  {Illingworth} G.,
  {Schechter} P.~L.,  1983, \mn@doi [\apj] {10.1086/160757}, \href
  {https://ui.adsabs.harvard.edu/abs/1983ApJ...266...41D} {266, 41}

\bibitem[\protect\citeauthoryear{{Deibel}, {Valluri}  \& {Merritt}}{{Deibel}
  et~al.}{2011}]{DVM11}
{Deibel} A.~T.,  {Valluri} M.,   {Merritt} D.,  2011, \mn@doi [\apj]
  {10.1088/0004-637X/728/2/128}, \href
  {https://ui.adsabs.harvard.edu/abs/2011ApJ...728..128D} {728, 128}

\bibitem[\protect\citeauthoryear{{Dimauro} et~al.,}{{Dimauro}
  et~al.}{2018}]{DHD18}
{Dimauro} P.,  et~al., 2018, \mn@doi [\mnras] {10.1093/mnras/sty1379}, \href
  {https://ui.adsabs.harvard.edu/abs/2018MNRAS.478.5410D} {478, 5410}

\bibitem[\protect\citeauthoryear{{Dom{\'\i}nguez S{\'a}nchez},
  {Huertas-Company}, {Bernardi}, {Tuccillo}  \& {Fischer}}{{Dom{\'\i}nguez
  S{\'a}nchez} et~al.}{2018}]{DHB18}
{Dom{\'\i}nguez S{\'a}nchez} H.,  {Huertas-Company} M.,  {Bernardi} M.,
  {Tuccillo} D.,   {Fischer} J.~L.,  2018, \mn@doi [\mnras]
  {10.1093/mnras/sty338}, \href
  {https://ui.adsabs.harvard.edu/abs/2018MNRAS.476.3661D} {476, 3661}

\bibitem[\protect\citeauthoryear{{Drory} et~al.,}{{Drory} et~al.}{2015}]{DMB15}
{Drory} N.,  et~al., 2015, \mn@doi [\aj] {10.1088/0004-6256/149/2/77}, \href
  {https://ui.adsabs.harvard.edu/abs/2015AJ....149...77D} {149, 77}

\bibitem[\protect\citeauthoryear{{Du}, {Ho}, {Debattista}, {Pillepich},
  {Nelson}, {Zhao}  \& {Hernquist}}{{Du} et~al.}{2020}]{DHD02}
{Du} M.,  {Ho} L.~C.,  {Debattista} V.~P.,  {Pillepich} A.,  {Nelson} D.,
  {Zhao} D.,   {Hernquist} L.,  2020, arXiv e-prints, \href
  {https://ui.adsabs.harvard.edu/abs/2020arXiv200204182D} {p. arXiv:2002.04182}

\bibitem[\protect\citeauthoryear{{Duckworth}, {Tojeiro}  \&
  {Kraljic}}{{Duckworth} et~al.}{2020}]{DTK20}
{Duckworth} C.,  {Tojeiro} R.,   {Kraljic} K.,  2020, \mn@doi [\mnras]
  {10.1093/mnras/stz3575}, \href
  {https://ui.adsabs.harvard.edu/abs/2020MNRAS.492.1869D} {492, 1869}

\bibitem[\protect\citeauthoryear{{Emsellem} et~al.,}{{Emsellem}
  et~al.}{2011}]{ECK11}
{Emsellem} E.,  et~al., 2011, \mn@doi [\mnras]
  {10.1111/j.1365-2966.2011.18496.x}, \href
  {https://ui.adsabs.harvard.edu/abs/2011MNRAS.414..888E} {414, 888}

\bibitem[\protect\citeauthoryear{{Erwin}}{{Erwin}}{2015}]{E15}
{Erwin} P.,  2015, \mn@doi [\apj] {10.1088/0004-637X/799/2/226}, \href
  {https://ui.adsabs.harvard.edu/abs/2015ApJ...799..226E} {799, 226}

\bibitem[\protect\citeauthoryear{{Faber} \& {Jackson}}{{Faber} \&
  {Jackson}}{1976}]{FJ76}
{Faber} S.~M.,  {Jackson} R.~E.,  1976, \mn@doi [\apj] {10.1086/154215}, \href
  {https://ui.adsabs.harvard.edu/abs/1976ApJ...204..668F} {204, 668}

\bibitem[\protect\citeauthoryear{{Fall} \& {Romanowsky}}{{Fall} \&
  {Romanowsky}}{2013}]{FR13}
{Fall} S.~M.,  {Romanowsky} A.~J.,  2013, \mn@doi [\apjl]
  {10.1088/2041-8205/769/2/L26}, \href
  {https://ui.adsabs.harvard.edu/abs/2013ApJ...769L..26F} {769, L26}

\bibitem[\protect\citeauthoryear{{Fall} \& {Romanowsky}}{{Fall} \&
  {Romanowsky}}{2018}]{FR18}
{Fall} S.~M.,  {Romanowsky} A.~J.,  2018, \mn@doi [\apj]
  {10.3847/1538-4357/aaeb27}, \href
  {https://ui.adsabs.harvard.edu/abs/2018ApJ...868..133F} {868, 133}

\bibitem[\protect\citeauthoryear{{Gadotti}}{{Gadotti}}{2009}]{G09}
{Gadotti} D.~A.,  2009, \mn@doi [\mnras] {10.1111/j.1365-2966.2008.14257.x},
  \href {https://ui.adsabs.harvard.edu/abs/2009MNRAS.393.1531G} {393, 1531}

\bibitem[\protect\citeauthoryear{{Gadotti} \& {Kauffmann}}{{Gadotti} \&
  {Kauffmann}}{2009}]{GK09}
{Gadotti} D.~A.,  {Kauffmann} G.,  2009, \mn@doi [\mnras]
  {10.1111/j.1365-2966.2009.15328.x}, \href
  {https://ui.adsabs.harvard.edu/abs/2009MNRAS.399..621G} {399, 621}

\bibitem[\protect\citeauthoryear{{Gargiulo} et~al.,}{{Gargiulo}
  et~al.}{2015}]{GCP15}
{Gargiulo} I.~D.,  et~al., 2015, \mn@doi [\mnras] {10.1093/mnras/stu2272},
  \href {https://ui.adsabs.harvard.edu/abs/2015MNRAS.446.3820G} {446, 3820}

\bibitem[\protect\citeauthoryear{{Gargiulo} et~al.,}{{Gargiulo}
  et~al.}{2019}]{GMG19}
{Gargiulo} I.~D.,  et~al., 2019, \mn@doi [\mnras] {10.1093/mnras/stz2536},
  \href {https://ui.adsabs.harvard.edu/abs/2019MNRAS.489.5742G} {489, 5742}

\bibitem[\protect\citeauthoryear{{Garrison-Kimmel} et~al.,}{{Garrison-Kimmel}
  et~al.}{2018}]{GHW18}
{Garrison-Kimmel} S.,  et~al., 2018, \mn@doi [\mnras] {10.1093/mnras/sty2513},
  \href {https://ui.adsabs.harvard.edu/abs/2018MNRAS.481.4133G} {481, 4133}

\bibitem[\protect\citeauthoryear{{Gorski}, {Wandelt}, {Hansen}, {Hivon}  \&
  {Banday}}{{Gorski} et~al.}{1999}]{GWH99}
{Gorski} K.~M.,  {Wandelt} B.~D.,  {Hansen} F.~K.,  {Hivon} E.,   {Banday}
  A.~J.,  1999, arXiv e-prints, \href
  {https://ui.adsabs.harvard.edu/abs/1999astro.ph..5275G} {pp
  astro--ph/9905275}

\bibitem[\protect\citeauthoryear{{G{\'o}rski}, {Banday}, {Hivon}  \& {Wand
  elt}}{{G{\'o}rski} et~al.}{2002}]{GBH02}
{G{\'o}rski} K.~M.,  {Banday} A.~J.,  {Hivon} E.,   {Wand elt} B.~D.,  2002, in
  Astronomical Data Analysis Software and Systems XI. p.~107

\bibitem[\protect\citeauthoryear{{Graham}}{{Graham}}{2012}]{G12}
{Graham} A.~W.,  2012, \mn@doi [\apj] {10.1088/0004-637X/746/1/113}, \href
  {https://ui.adsabs.harvard.edu/abs/2012ApJ...746..113G} {746, 113}

\bibitem[\protect\citeauthoryear{{Grand} et~al.,}{{Grand} et~al.}{2017}]{GGM17}
{Grand} R. J.~J.,  et~al., 2017, \mn@doi [\mnras] {10.1093/mnras/stx071}, \href
  {https://ui.adsabs.harvard.edu/abs/2017MNRAS.467..179G} {467, 179}

\bibitem[\protect\citeauthoryear{{Greene}, {Ho}  \& {Barth}}{{Greene}
  et~al.}{2008}]{GHB08}
{Greene} J.~E.,  {Ho} L.~C.,   {Barth} A.~J.,  2008, \mn@doi [\apj]
  {10.1086/592078}, \href
  {https://ui.adsabs.harvard.edu/abs/2008ApJ...688..159G} {688, 159}

\bibitem[\protect\citeauthoryear{{Habouzit} et~al.,}{{Habouzit}
  et~al.}{2020}]{HLS20}
{Habouzit} M.,  et~al., 2020, arXiv e-prints, \href
  {https://ui.adsabs.harvard.edu/abs/2020arXiv200610094H} {p. arXiv:2006.10094}

\bibitem[\protect\citeauthoryear{{Hagberg}, {Schult}  \& {Swart}}{{Hagberg}
  et~al.}{2008}]{HSS08}
{Hagberg} A.~A.,  {Schult} D.~A.,   {Swart} P.~J.,  2008, in Proceedings of the
  7th Python in Science Conference. Pasadena, CA USA, pp 11 -- 15, \url
  {http://conference.scipy.org/proceedings/SciPy2008/paper_2/}

\bibitem[\protect\citeauthoryear{{H{\"a}ring} \& {Rix}}{{H{\"a}ring} \&
  {Rix}}{2004}]{HR04}
{H{\"a}ring} N.,  {Rix} H.-W.,  2004, \mn@doi [\apjl] {10.1086/383567}, \href
  {https://ui.adsabs.harvard.edu/abs/2004ApJ...604L..89H} {604, L89}

\bibitem[\protect\citeauthoryear{{Hopkins}, {Cox}, {Younger}  \&
  {Hernquist}}{{Hopkins} et~al.}{2009}]{HCY08}
{Hopkins} P.~F.,  {Cox} T.~J.,  {Younger} J.~D.,   {Hernquist} L.,  2009,
  \mn@doi [\apj] {10.1088/0004-637X/691/2/1168}, \href
  {https://ui.adsabs.harvard.edu/abs/2009ApJ...691.1168H} {691, 1168}

\bibitem[\protect\citeauthoryear{{Hopkins} et~al.,}{{Hopkins}
  et~al.}{2010}]{HBC10}
{Hopkins} P.~F.,  et~al., 2010, \mn@doi [\apj] {10.1088/0004-637X/715/1/202},
  \href {https://ui.adsabs.harvard.edu/abs/2010ApJ...715..202H} {715, 202}

\bibitem[\protect\citeauthoryear{{Hunter}}{{Hunter}}{2007}]{H07}
{Hunter} J.~D.,  2007, \mn@doi [Computing in Science and Engineering]
  {10.1109/MCSE.2007.55}, \href
  {https://ui.adsabs.harvard.edu/abs/2007CSE.....9...90H} {9, 90}

\bibitem[\protect\citeauthoryear{{Hunter} et~al.,}{{Hunter}
  et~al.}{2012}]{HFA12}
{Hunter} D.~A.,  et~al., 2012, \mn@doi [\aj] {10.1088/0004-6256/144/5/134},
  \href {https://ui.adsabs.harvard.edu/abs/2012AJ....144..134H} {144, 134}

\bibitem[\protect\citeauthoryear{{Irodotou}, {Thomas}, {Henriques}, {Sargent}
  \& {Hislop}}{{Irodotou} et~al.}{2019}]{ITH19}
{Irodotou} D.,  {Thomas} P.~A.,  {Henriques} B.~M.,  {Sargent} M.~T.,
  {Hislop} J.~M.,  2019, \mn@doi [\mnras] {10.1093/mnras/stz2365}, \href
  {https://ui.adsabs.harvard.edu/abs/2019MNRAS.489.3609I} {489, 3609}

\bibitem[\protect\citeauthoryear{{Jackson}, {Martin}, {Kaviraj}, {Laigle},
  {Devriendt}, {Dubois}  \& {Pichon}}{{Jackson} et~al.}{2020}]{JMK20}
{Jackson} R.~A.,  {Martin} G.,  {Kaviraj} S.,  {Laigle} C.,  {Devriendt}
  J.~E.~G.,  {Dubois} Y.,   {Pichon} C.,  2020, \mn@doi [\mnras]
  {10.1093/mnras/staa970}, \href
  {https://ui.adsabs.harvard.edu/abs/2020MNRAS.494.5568J} {494, 5568}

\bibitem[\protect\citeauthoryear{{Joshi}, {Pillepich}, {Nelson}, {Marinacci},
  {Springel}, {Rodriguez-Gomez}, {Vogelsberger}  \& {Hernquist}}{{Joshi}
  et~al.}{2020}]{JPN20}
{Joshi} G.~D.,  {Pillepich} A.,  {Nelson} D.,  {Marinacci} F.,  {Springel} V.,
  {Rodriguez-Gomez} V.,  {Vogelsberger} M.,   {Hernquist} L.,  2020, arXiv
  e-prints, \href {https://ui.adsabs.harvard.edu/abs/2020arXiv200401191J} {p.
  arXiv:2004.01191}

\bibitem[\protect\citeauthoryear{{Kent}}{{Kent}}{1986}]{K86}
{Kent} S.~M.,  1986, \mn@doi [\aj] {10.1086/114106}, \href
  {http://adsabs.harvard.edu/abs/1986AJ.....91.1301K} {91, 1301}

\bibitem[\protect\citeauthoryear{{Kent}}{{Kent}}{1987}]{K87}
{Kent} S.~M.,  1987, \mn@doi [\aj] {10.1086/114366}, \href
  {http://adsabs.harvard.edu/abs/1987AJ.....93..816K} {93, 816}

\bibitem[\protect\citeauthoryear{{Kent}}{{Kent}}{1988}]{K88}
{Kent} S.~M.,  1988, \mn@doi [\aj] {10.1086/114829}, \href
  {http://adsabs.harvard.edu/abs/1988AJ.....96..514K} {96, 514}

\bibitem[\protect\citeauthoryear{{Koribalski} et~al.,}{{Koribalski}
  et~al.}{2018}]{KWK18}
{Koribalski} B.~S.,  et~al., 2018, \mn@doi [\mnras] {10.1093/mnras/sty479},
  \href {https://ui.adsabs.harvard.edu/abs/2018MNRAS.478.1611K} {478, 1611}

\bibitem[\protect\citeauthoryear{{Kormendy}}{{Kormendy}}{2016}]{K16}
{Kormendy} J.,  2016, {Elliptical Galaxies and Bulges of Disc Galaxies: Summary
  of Progress and Outstanding Issues}.
p.~431, \mn@doi{10.1007/978-3-319-19378-6_16}

\bibitem[\protect\citeauthoryear{{Kormendy} \& {Gebhardt}}{{Kormendy} \&
  {Gebhardt}}{2001}]{KG01}
{Kormendy} J.,  {Gebhardt} K.,  2001, in {Wheeler} J.~C.,  {Martel} H.,  eds,
  American Institute of Physics Conference Series Vol. 586, 20th Texas
  Symposium on relativistic astrophysics. pp 363--381 (\mn@eprint {arXiv}
  {astro-ph/0105230}), \mn@doi{10.1063/1.1419581}

\bibitem[\protect\citeauthoryear{{Kormendy} \& {Illingworth}}{{Kormendy} \&
  {Illingworth}}{1982}]{KI82}
{Kormendy} J.,  {Illingworth} G.,  1982, \mn@doi [\apj] {10.1086/159923}, \href
  {https://ui.adsabs.harvard.edu/abs/1982ApJ...256..460K} {256, 460}

\bibitem[\protect\citeauthoryear{{Kormendy} \& {Kennicutt}}{{Kormendy} \&
  {Kennicutt}}{2004}]{KK04}
{Kormendy} J.,  {Kennicutt} Robert~C. J.,  2004, \mn@doi [\araa]
  {10.1146/annurev.astro.42.053102.134024}, \href
  {https://ui.adsabs.harvard.edu/abs/2004ARA&A..42..603K} {42, 603}

\bibitem[\protect\citeauthoryear{{Krajnovi{\'c}} et~al.,}{{Krajnovi{\'c}}
  et~al.}{2015}]{KWU15}
{Krajnovi{\'c}} D.,  et~al., 2015, \mn@doi [\mnras] {10.1093/mnras/stv958},
  \href {https://ui.adsabs.harvard.edu/abs/2015MNRAS.452....2K} {452, 2}

\bibitem[\protect\citeauthoryear{{Lagos}, {Cora}  \& {Padilla}}{{Lagos}
  et~al.}{2008}]{LCP08}
{Lagos} C. D.~P.,  {Cora} S.~A.,   {Padilla} N.~D.,  2008, \mn@doi [\mnras]
  {10.1111/j.1365-2966.2008.13456.x}, \href
  {https://ui.adsabs.harvard.edu/abs/2008MNRAS.388..587L} {388, 587}

\bibitem[\protect\citeauthoryear{{Lagos} et~al.,}{{Lagos}
  et~al.}{2018a}]{LSB18a}
{Lagos} C. d.~P.,  et~al., 2018a, \mn@doi [\mnras] {10.1093/mnras/stx2667},
  \href {https://ui.adsabs.harvard.edu/abs/2018MNRAS.473.4956L} {473, 4956}

\bibitem[\protect\citeauthoryear{{Lagos}, {Schaye}, {Bah{\'e}}, {Van de Sande},
  {Kay}, {Barnes}, {Davis}  \& {Dalla Vecchia}}{{Lagos} et~al.}{2018b}]{LSB18b}
{Lagos} C. d.~P.,  {Schaye} J.,  {Bah{\'e}} Y.,  {Van de Sande} J.,  {Kay}
  S.~T.,  {Barnes} D.,  {Davis} T.~A.,   {Dalla Vecchia} C.,  2018b, \mn@doi
  [\mnras] {10.1093/mnras/sty489}, \href
  {https://ui.adsabs.harvard.edu/abs/2018MNRAS.476.4327L} {476, 4327}

\bibitem[\protect\citeauthoryear{{Lelli}, {McGaugh}  \& {Schombert}}{{Lelli}
  et~al.}{2016}]{LMS16}
{Lelli} F.,  {McGaugh} S.~S.,   {Schombert} J.~M.,  2016, \mn@doi [\aj]
  {10.3847/0004-6256/152/6/157}, \href
  {https://ui.adsabs.harvard.edu/abs/2016AJ....152..157L} {152, 157}

\bibitem[\protect\citeauthoryear{{Lingard} et~al.,}{{Lingard}
  et~al.}{2020}]{LMK20}
{Lingard} T.~K.,  et~al., 2020, arXiv e-prints, \href
  {https://ui.adsabs.harvard.edu/abs/2020arXiv200610450L} {p. arXiv:2006.10450}

\bibitem[\protect\citeauthoryear{{MacArthur}, {Courteau}  \&
  {Holtzman}}{{MacArthur} et~al.}{2003}]{MCH03}
{MacArthur} L.~A.,  {Courteau} S.,   {Holtzman} J.~A.,  2003, \mn@doi [\apj]
  {10.1086/344506}, \href
  {https://ui.adsabs.harvard.edu/abs/2003ApJ...582..689M} {582, 689}

\bibitem[\protect\citeauthoryear{{Mancera Pi{\~n}a}, {Posti}, {Fraternali},
  {Adams}  \& {Oosterloo}}{{Mancera Pi{\~n}a} et~al.}{2020}]{MPPF20}
{Mancera Pi{\~n}a} P.~E.,  {Posti} L.,  {Fraternali} F.,  {Adams} E. A.~K.,
  {Oosterloo} T.,  2020, arXiv e-prints, \href
  {https://ui.adsabs.harvard.edu/abs/2020arXiv200906645M} {p. arXiv:2009.06645}

\bibitem[\protect\citeauthoryear{{Mancini} et~al.,}{{Mancini}
  et~al.}{2019}]{MDJ19}
{Mancini} C.,  et~al., 2019, \mn@doi [\mnras] {10.1093/mnras/stz2130}, \href
  {https://ui.adsabs.harvard.edu/abs/2019MNRAS.489.1265M} {489, 1265}

\bibitem[\protect\citeauthoryear{{Marconi} \& {Hunt}}{{Marconi} \&
  {Hunt}}{2003}]{MH03}
{Marconi} A.,  {Hunt} L.~K.,  2003, \mn@doi [\apjl] {10.1086/375804}, \href
  {https://ui.adsabs.harvard.edu/abs/2003ApJ...589L..21M} {589, L21}

\bibitem[\protect\citeauthoryear{{Marinacci}, {Pakmor}  \&
  {Springel}}{{Marinacci} et~al.}{2014}]{MPS14}
{Marinacci} F.,  {Pakmor} R.,   {Springel} V.,  2014, \mn@doi [\mnras]
  {10.1093/mnras/stt2003}, \href
  {https://ui.adsabs.harvard.edu/abs/2014MNRAS.437.1750M} {437, 1750}

\bibitem[\protect\citeauthoryear{{Masters} \& {Galaxy Zoo Team}}{{Masters} \&
  {Galaxy Zoo Team}}{2020}]{MG20}
{Masters} K.~L.,  {Galaxy Zoo Team} 2020, in {Valluri} M.,  {Sellwood} J.~A.,
  eds,  IAU Symposium Vol. 353, IAU Symposium. pp 205--212 (\mn@eprint {arXiv}
  {1910.08177}), \mn@doi{10.1017/S1743921319008615}

\bibitem[\protect\citeauthoryear{{McAlpine} et~al.,}{{McAlpine}
  et~al.}{2016}]{MHS16}
{McAlpine} S.,  et~al., 2016, \mn@doi [Astronomy and Computing]
  {10.1016/j.ascom.2016.02.004}, \href
  {https://ui.adsabs.harvard.edu/abs/2016A&C....15...72M} {15, 72}

\bibitem[\protect\citeauthoryear{{McConnell} \& {Ma}}{{McConnell} \&
  {Ma}}{2013}]{MM13}
{McConnell} N.~J.,  {Ma} C.-P.,  2013, \mn@doi [\apj]
  {10.1088/0004-637X/764/2/184}, \href
  {https://ui.adsabs.harvard.edu/abs/2013ApJ...764..184M} {764, 184}

\bibitem[\protect\citeauthoryear{{McGaugh}, {Schombert}, {Bothun}  \& {de
  Blok}}{{McGaugh} et~al.}{2000}]{MSB00}
{McGaugh} S.~S.,  {Schombert} J.~M.,  {Bothun} G.~D.,   {de Blok} W.~J.~G.,
  2000, \mn@doi [\apjl] {10.1086/312628}, \href
  {http://adsabs.harvard.edu/abs/2000ApJ...533L..99M} {533, L99}

\bibitem[\protect\citeauthoryear{{{M}c{K}inney}}{{{M}c{K}inney}}{2010}]{M10}
{{M}c{K}inney} W.,  2010, in {P}roceedings of the 9th {P}ython in {S}cience
  {C}onference. pp 56 -- 61, \mn@doi{10.25080/Majora-92bf1922-00a}

\bibitem[\protect\citeauthoryear{{Moffett} et~al.,}{{Moffett}
  et~al.}{2016}]{MLD16}
{Moffett} A.~J.,  et~al., 2016, \mn@doi [\mnras] {10.1093/mnras/stw1861}, \href
  {https://ui.adsabs.harvard.edu/abs/2016MNRAS.462.4336M} {462, 4336}

\bibitem[\protect\citeauthoryear{{Monachesi} et~al.,}{{Monachesi}
  et~al.}{2019}]{MGG19}
{Monachesi} A.,  et~al., 2019, \mn@doi [\mnras] {10.1093/mnras/stz538}, \href
  {https://ui.adsabs.harvard.edu/abs/2019MNRAS.485.2589M} {485, 2589}

\bibitem[\protect\citeauthoryear{{Naab} et~al.,}{{Naab} et~al.}{2014}]{NOE14}
{Naab} T.,  et~al., 2014, \mn@doi [\mnras] {10.1093/mnras/stt1919}, \href
  {https://ui.adsabs.harvard.edu/abs/2014MNRAS.444.3357N} {444, 3357}

\bibitem[\protect\citeauthoryear{{Obreja}, {Dom{\'\i}nguez-Tenreiro}, {Brook},
  {Mart{\'\i}nez-Serrano}, {Dom{\'e}nech-Moral}, {Serna}, {Moll{\'a}}  \&
  {Stinson}}{{Obreja} et~al.}{2013}]{ODB13}
{Obreja} A.,  {Dom{\'\i}nguez-Tenreiro} R.,  {Brook} C.,
  {Mart{\'\i}nez-Serrano} F.~J.,  {Dom{\'e}nech-Moral} M.,  {Serna} A.,
  {Moll{\'a}} M.,   {Stinson} G.,  2013, \mn@doi [\apj]
  {10.1088/0004-637X/763/1/26}, \href
  {https://ui.adsabs.harvard.edu/abs/2013ApJ...763...26O} {763, 26}

\bibitem[\protect\citeauthoryear{{Obreja}, {Macci{\`o}}, {Moster}, {Dutton},
  {Buck}, {Stinson}  \& {Wang}}{{Obreja} et~al.}{2018}]{OMM18}
{Obreja} A.,  {Macci{\`o}} A.~V.,  {Moster} B.,  {Dutton} A.~A.,  {Buck} T.,
  {Stinson} G.~S.,   {Wang} L.,  2018, \mn@doi [\mnras]
  {10.1093/mnras/sty1022}, \href
  {https://ui.adsabs.harvard.edu/abs/2018MNRAS.477.4915O} {477, 4915}

\bibitem[\protect\citeauthoryear{{Obreschkow} \& {Glazebrook}}{{Obreschkow} \&
  {Glazebrook}}{2014}]{OG14}
{Obreschkow} D.,  {Glazebrook} K.,  2014, \mn@doi [\apj]
  {10.1088/0004-637X/784/1/26}, \href
  {https://ui.adsabs.harvard.edu/abs/2014ApJ...784...26O} {784, 26}

\bibitem[\protect\citeauthoryear{{Obreschkow}, {Elahi}, {Lagos}, {Poulton}  \&
  {Ludlow}}{{Obreschkow} et~al.}{2020}]{OEL20}
{Obreschkow} D.,  {Elahi} P.~J.,  {Lagos} C. d.~P.,  {Poulton} R. J.~J.,
  {Ludlow} A.~D.,  2020, \mn@doi [\mnras] {10.1093/mnras/staa445}, \href
  {https://ui.adsabs.harvard.edu/abs/2020MNRAS.493.4551O} {493, 4551}

\bibitem[\protect\citeauthoryear{{Oh} et~al.,}{{Oh} et~al.}{2020}]{OCB20}
{Oh} S.,  et~al., 2020, \mn@doi [\mnras] {10.1093/mnras/staa1330}, \href
  {https://ui.adsabs.harvard.edu/abs/2020MNRAS.495.4638O} {495, 4638}

\bibitem[\protect\citeauthoryear{{Ott} et~al.,}{{Ott} et~al.}{2012}]{OSW12}
{Ott} J.,  et~al., 2012, \mn@doi [\aj] {10.1088/0004-6256/144/4/123}, \href
  {https://ui.adsabs.harvard.edu/abs/2012AJ....144..123O} {144, 123}

\bibitem[\protect\citeauthoryear{{Park} et~al.,}{{Park} et~al.}{2019}]{PYD19}
{Park} M.-J.,  et~al., 2019, \mn@doi [\apj] {10.3847/1538-4357/ab3afe}, \href
  {https://ui.adsabs.harvard.edu/abs/2019ApJ...883...25P} {883, 25}

\bibitem[\protect\citeauthoryear{{Paturel}, {Petit}, {Prugniel}, {Theureau},
  {Rousseau}, {Brouty}, {Dubois}  \& {Cambr{\'e}sy}}{{Paturel}
  et~al.}{2003}]{PPP03}
{Paturel} G.,  {Petit} C.,  {Prugniel} P.,  {Theureau} G.,  {Rousseau} J.,
  {Brouty} M.,  {Dubois} P.,   {Cambr{\'e}sy} L.,  2003, \mn@doi [\aap]
  {10.1051/0004-6361:20031411}, \href
  {https://ui.adsabs.harvard.edu/abs/2003A&A...412...45P} {412, 45}

\bibitem[\protect\citeauthoryear{{Pedrosa} \& {Tissera}}{{Pedrosa} \&
  {Tissera}}{2015}]{PT15}
{Pedrosa} S.~E.,  {Tissera} P.~B.,  2015, \mn@doi [\aap]
  {10.1051/0004-6361/201526440}, \href
  {https://ui.adsabs.harvard.edu/abs/2015A&A...584A..43P} {584, A43}

\bibitem[\protect\citeauthoryear{{Peebles}}{{Peebles}}{2020}]{P20}
{Peebles} P.~J.~E.,  2020, arXiv e-prints, \href
  {https://ui.adsabs.harvard.edu/abs/2020arXiv200507588P} {p. arXiv:2005.07588}

\bibitem[\protect\citeauthoryear{{Peng}, {Ho}, {Impey}  \& {Rix}}{{Peng}
  et~al.}{2002}]{PHI02}
{Peng} C.~Y.,  {Ho} L.~C.,  {Impey} C.~D.,   {Rix} H.-W.,  2002, \mn@doi [\aj]
  {10.1086/340952}, \href
  {https://ui.adsabs.harvard.edu/abs/2002AJ....124..266P} {124, 266}

\bibitem[\protect\citeauthoryear{{Pillepich}, {Madau}  \& {Mayer}}{{Pillepich}
  et~al.}{2015}]{PMM15}
{Pillepich} A.,  {Madau} P.,   {Mayer} L.,  2015, \mn@doi [\apj]
  {10.1088/0004-637X/799/2/184}, \href
  {https://ui.adsabs.harvard.edu/abs/2015ApJ...799..184P} {799, 184}

\bibitem[\protect\citeauthoryear{{Ponomareva}, {Verheijen}  \&
  {Bosma}}{{Ponomareva} et~al.}{2016}]{PVB16}
{Ponomareva} A.~A.,  {Verheijen} M. A.~W.,   {Bosma} A.,  2016, \mn@doi
  [\mnras] {10.1093/mnras/stw2213}, \href
  {https://ui.adsabs.harvard.edu/abs/2016MNRAS.463.4052P} {463, 4052}

\bibitem[\protect\citeauthoryear{{Posti}, {Fraternali}, {Di Teodoro}  \&
  {Pezzulli}}{{Posti} et~al.}{2018}]{PFD18}
{Posti} L.,  {Fraternali} F.,  {Di Teodoro} E.~M.,   {Pezzulli} G.,  2018,
  \mn@doi [\aap] {10.1051/0004-6361/201833091}, \href
  {https://ui.adsabs.harvard.edu/abs/2018A&A...612L...6P} {612, L6}

\bibitem[\protect\citeauthoryear{{Prugniel} \& {Simien}}{{Prugniel} \&
  {Simien}}{1997}]{PS97}
{Prugniel} P.,  {Simien} F.,  1997, \aap, \href
  {https://ui.adsabs.harvard.edu/abs/1997A&A...321..111P} {321, 111}

\bibitem[\protect\citeauthoryear{{Romanowsky} \& {Fall}}{{Romanowsky} \&
  {Fall}}{2012}]{RF12}
{Romanowsky} A.~J.,  {Fall} S.~M.,  2012, \mn@doi [\apjs]
  {10.1088/0067-0049/203/2/17}, \href
  {https://ui.adsabs.harvard.edu/abs/2012ApJS..203...17R} {203, 17}

\bibitem[\protect\citeauthoryear{{Rosas-Guevara} et~al.,}{{Rosas-Guevara}
  et~al.}{2020}]{R20}
{Rosas-Guevara} Y.,  et~al., 2020, \mn@doi [\mnras] {10.1093/mnras/stz3180},
  \href {https://ui.adsabs.harvard.edu/abs/2020MNRAS.491.2547R} {491, 2547}

\bibitem[\protect\citeauthoryear{{Rosito}, {Pedrosa}, {Tissera}, {Avila-Reese},
  {Lacerna}, {Bignone}, {Ibarra-Medel}  \& {Varela}}{{Rosito}
  et~al.}{2018}]{RPT18}
{Rosito} M.~S.,  {Pedrosa} S.~E.,  {Tissera} P.~B.,  {Avila-Reese} V.,
  {Lacerna} I.,  {Bignone} L.~A.,  {Ibarra-Medel} H.~J.,   {Varela} S.,  2018,
  \mn@doi [\aap] {10.1051/0004-6361/201732302}, \href
  {https://ui.adsabs.harvard.edu/abs/2018A&A...614A..85R} {614, A85}

\bibitem[\protect\citeauthoryear{{Rosito}, {Tissera}, {Pedrosa}  \&
  {Rosas-Guevara}}{{Rosito} et~al.}{2019}]{RTP19}
{Rosito} M.~S.,  {Tissera} P.~B.,  {Pedrosa} S.~E.,   {Rosas-Guevara} Y.,
  2019, \mn@doi [\aap] {10.1051/0004-6361/201834720}, \href
  {https://ui.adsabs.harvard.edu/abs/2019A&A...629A..37R} {629, A37}

\bibitem[\protect\citeauthoryear{{Sales}, {Navarro}, {Theuns}, {Schaye},
  {White}, {Frenk}, {Crain}  \& {Dalla Vecchia}}{{Sales} et~al.}{2012}]{SNT12}
{Sales} L.~V.,  {Navarro} J.~F.,  {Theuns} T.,  {Schaye} J.,  {White} S. D.~M.,
   {Frenk} C.~S.,  {Crain} R.~A.,   {Dalla Vecchia} C.,  2012, \mn@doi [\mnras]
  {10.1111/j.1365-2966.2012.20975.x}, \href
  {https://ui.adsabs.harvard.edu/abs/2012MNRAS.423.1544S} {423, 1544}

\bibitem[\protect\citeauthoryear{{S{\'a}nchez} et~al.,}{{S{\'a}nchez}
  et~al.}{2012}]{SKG12}
{S{\'a}nchez} S.~F.,  et~al., 2012, \mn@doi [\aap]
  {10.1051/0004-6361/201117353}, \href
  {https://ui.adsabs.harvard.edu/abs/2012A&A...538A...8S} {538, A8}

\bibitem[\protect\citeauthoryear{{Sani}, {Marconi}, {Hunt}  \&
  {Risaliti}}{{Sani} et~al.}{2011}]{SMH11}
{Sani} E.,  {Marconi} A.,  {Hunt} L.~K.,   {Risaliti} G.,  2011, \mn@doi
  [\mnras] {10.1111/j.1365-2966.2011.18229.x}, \href
  {https://ui.adsabs.harvard.edu/abs/2011MNRAS.413.1479S} {413, 1479}

\bibitem[\protect\citeauthoryear{{Scannapieco}, {White}, {Springel}  \&
  {Tissera}}{{Scannapieco} et~al.}{2009}]{SWS09}
{Scannapieco} C.,  {White} S. D.~M.,  {Springel} V.,   {Tissera} P.~B.,  2009,
  \mn@doi [\mnras] {10.1111/j.1365-2966.2009.14764.x}, \href
  {https://ui.adsabs.harvard.edu/abs/2009MNRAS.396..696S} {396, 696}

\bibitem[\protect\citeauthoryear{{Schaye} et~al.,}{{Schaye}
  et~al.}{2015}]{SCB15}
{Schaye} J.,  et~al., 2015, \mn@doi [\mnras] {10.1093/mnras/stu2058}, \href
  {https://ui.adsabs.harvard.edu/abs/2015MNRAS.446..521S} {446, 521}

\bibitem[\protect\citeauthoryear{{Scorza} \& {Bender}}{{Scorza} \&
  {Bender}}{1995}]{SB95}
{Scorza} C.,  {Bender} R.,  1995, \aap, \href
  {https://ui.adsabs.harvard.edu/abs/1995A&A...293...20S} {293, 20}

\bibitem[\protect\citeauthoryear{{Sersic}}{{Sersic}}{1968}]{S68}
{Sersic} J.~L.,  1968, {Atlas de Galaxias Australes}

\bibitem[\protect\citeauthoryear{{Shankar}, {Marulli}, {Mathur}, {Bernardi}  \&
  {Bournaud}}{{Shankar} et~al.}{2012}]{SMM12}
{Shankar} F.,  {Marulli} F.,  {Mathur} S.,  {Bernardi} M.,   {Bournaud} F.,
  2012, \mn@doi [\aap] {10.1051/0004-6361/201118387}, \href
  {https://ui.adsabs.harvard.edu/abs/2012A&A...540A..23S} {540, A23}

\bibitem[\protect\citeauthoryear{{Simard} et~al.,}{{Simard}
  et~al.}{2002}]{SWV02}
{Simard} L.,  et~al., 2002, \mn@doi [\apjs] {10.1086/341399}, \href
  {https://ui.adsabs.harvard.edu/abs/2002ApJS..142....1S} {142, 1}

\bibitem[\protect\citeauthoryear{{Springel}, {White}, {Tormen}  \&
  {Kauffmann}}{{Springel} et~al.}{2001}]{SWT01}
{Springel} V.,  {White} S. D.~M.,  {Tormen} G.,   {Kauffmann} G.,  2001,
  \mn@doi [\mnras] {10.1046/j.1365-8711.2001.04912.x}, \href
  {https://ui.adsabs.harvard.edu/abs/2001MNRAS.328..726S} {328, 726}

\bibitem[\protect\citeauthoryear{{Sweet}, {Fisher}, {Glazebrook}, {Obreschkow},
  {Lagos}  \& {Wang}}{{Sweet} et~al.}{2018}]{SFG18}
{Sweet} S.~M.,  {Fisher} D.,  {Glazebrook} K.,  {Obreschkow} D.,  {Lagos} C.,
  {Wang} L.,  2018, \mn@doi [\apj] {10.3847/1538-4357/aabfc4}, \href
  {https://ui.adsabs.harvard.edu/abs/2018ApJ...860...37S} {860, 37}

\bibitem[\protect\citeauthoryear{{Tabor}, {Merrifield}, {Arag{\'o}n-Salamanca},
  {Fraser-McKelvie}, {Peterken}, {Smethurst}, {Drory}  \& {Lane}}{{Tabor}
  et~al.}{2019}]{TMA19}
{Tabor} M.,  {Merrifield} M.,  {Arag{\'o}n-Salamanca} A.,  {Fraser-McKelvie}
  A.,  {Peterken} T.,  {Smethurst} R.,  {Drory} N.,   {Lane} R.~R.,  2019,
  \mn@doi [\mnras] {10.1093/mnras/stz431}, \href
  {https://ui.adsabs.harvard.edu/abs/2019MNRAS.485.1546T} {485, 1546}

\bibitem[\protect\citeauthoryear{{Tacchella} et~al.,}{{Tacchella}
  et~al.}{2019}]{TDH19}
{Tacchella} S.,  et~al., 2019, \mn@doi [\mnras] {10.1093/mnras/stz1657}, \href
  {https://ui.adsabs.harvard.edu/abs/2019MNRAS.487.5416T} {487, 5416}

\bibitem[\protect\citeauthoryear{{Thanjavur}, {Simard}, {Bluck}  \&
  {Mendel}}{{Thanjavur} et~al.}{2016}]{TSB16}
{Thanjavur} K.,  {Simard} L.,  {Bluck} A. F.~L.,   {Mendel} T.,  2016, \mn@doi
  [\mnras] {10.1093/mnras/stw495}, \href
  {https://ui.adsabs.harvard.edu/abs/2016MNRAS.459...44T} {459, 44}

\bibitem[\protect\citeauthoryear{{The EAGLE team}}{{The EAGLE
  team}}{2017}]{TET17}
{The EAGLE team} 2017, arXiv e-prints, \href
  {https://ui.adsabs.harvard.edu/abs/2017arXiv170609899T} {p. arXiv:1706.09899}

\bibitem[\protect\citeauthoryear{{Thob} et~al.,}{{Thob} et~al.}{2019}]{TCM19}
{Thob} A. C.~R.,  et~al., 2019, \mn@doi [\mnras] {10.1093/mnras/stz448}, \href
  {https://ui.adsabs.harvard.edu/abs/2019MNRAS.485..972T} {485, 972}

\bibitem[\protect\citeauthoryear{{Tissera}, {White}  \&
  {Scannapieco}}{{Tissera} et~al.}{2012}]{TWS12}
{Tissera} P.~B.,  {White} S. D.~M.,   {Scannapieco} C.,  2012, \mn@doi [\mnras]
  {10.1111/j.1365-2966.2011.20028.x}, \href
  {https://ui.adsabs.harvard.edu/abs/2012MNRAS.420..255T} {420, 255}

\bibitem[\protect\citeauthoryear{{Tissera}, {Rosas-Guevara}, {Bower}, {Crain},
  {del P Lagos}, {Schaller}, {Schaye}  \& {Theuns}}{{Tissera}
  et~al.}{2019}]{TRB19}
{Tissera} P.~B.,  {Rosas-Guevara} Y.,  {Bower} R.~G.,  {Crain} R.~A.,  {del P
  Lagos} C.,  {Schaller} M.,  {Schaye} J.,   {Theuns} T.,  2019, \mn@doi
  [\mnras] {10.1093/mnras/sty2817}, \href
  {https://ui.adsabs.harvard.edu/abs/2019MNRAS.482.2208T} {482, 2208}

\bibitem[\protect\citeauthoryear{{Toomre}}{{Toomre}}{1977}]{T77}
{Toomre} A.,  1977, in {Tinsley} B.~M.,  {Larson} Richard B.~Gehret D.~C.,
  eds, Evolution of Galaxies and Stellar Populations. p.~401

\bibitem[\protect\citeauthoryear{{Trayford}, {Frenk}, {Theuns}, {Schaye}  \&
  {Correa}}{{Trayford} et~al.}{2019}]{TFT19}
{Trayford} J.~W.,  {Frenk} C.~S.,  {Theuns} T.,  {Schaye} J.,   {Correa} C.,
  2019, \mn@doi [\mnras] {10.1093/mnras/sty2860}, \href
  {https://ui.adsabs.harvard.edu/abs/2019MNRAS.483..744T} {483, 744}

\bibitem[\protect\citeauthoryear{{Trussler}, {Maiolino}, {Maraston}, {Peng},
  {Thomas}, {Goddard}  \& {Lian}}{{Trussler} et~al.}{2020}]{TMM20}
{Trussler} J.,  {Maiolino} R.,  {Maraston} C.,  {Peng} Y.,  {Thomas} D.,
  {Goddard} D.,   {Lian} J.,  2020, arXiv e-prints, \href
  {https://ui.adsabs.harvard.edu/abs/2020arXiv200601154T} {p. arXiv:2006.01154}

\bibitem[\protect\citeauthoryear{{Tully} \& {Fisher}}{{Tully} \&
  {Fisher}}{1977}]{TF77}
{Tully} R.~B.,  {Fisher} J.~R.,  1977, \aap, \href
  {http://adsabs.harvard.edu/abs/1977A%26A....54..661T} {54, 661}

\bibitem[\protect\citeauthoryear{{Valluri} \& {Merritt}}{{Valluri} \&
  {Merritt}}{1998}]{VM98}
{Valluri} M.,  {Merritt} D.,  1998, \mn@doi [\apj] {10.1086/306269}, \href
  {https://ui.adsabs.harvard.edu/abs/1998ApJ...506..686V} {506, 686}

\bibitem[\protect\citeauthoryear{{Vergani}, {Pizzella}, {Corsini}, {van Driel},
  {Buson}, {Dettmar}  \& {Bertola}}{{Vergani} et~al.}{2007}]{VPC07}
{Vergani} D.,  {Pizzella} A.,  {Corsini} E.~M.,  {van Driel} W.,  {Buson}
  L.~M.,  {Dettmar} R.~J.,   {Bertola} F.,  2007, \mn@doi [\aap]
  {10.1051/0004-6361:20066413}, \href
  {https://ui.adsabs.harvard.edu/abs/2007A&A...463..883V} {463, 883}

\bibitem[\protect\citeauthoryear{{Virtanen} et~al.,}{{Virtanen}
  et~al.}{2020}]{VGO20}
{Virtanen} P.,  et~al., 2020, \mn@doi [Nature Methods]
  {10.1038/s41592-019-0686-2}, \href
  {https://ui.adsabs.harvard.edu/abs/2020NatMe..17..261V} {17, 261}

\bibitem[\protect\citeauthoryear{{Walter}, {Brinks}, {de Blok}, {Bigiel},
  {Kennicutt}, {Thornley}  \& {Leroy}}{{Walter} et~al.}{2008}]{WBD08}
{Walter} F.,  {Brinks} E.,  {de Blok} W.~J.~G.,  {Bigiel} F.,  {Kennicutt}
  Robert~C. J.,  {Thornley} M.~D.,   {Leroy} A.,  2008, \mn@doi [\aj]
  {10.1088/0004-6256/136/6/2563}, \href
  {https://ui.adsabs.harvard.edu/abs/2008AJ....136.2563W} {136, 2563}

\bibitem[\protect\citeauthoryear{{Wang} et~al.,}{{Wang} et~al.}{2019}]{WOL19}
{Wang} L.,  et~al., 2019, \mn@doi [\mnras] {10.1093/mnras/sty3010}, \href
  {https://ui.adsabs.harvard.edu/abs/2019MNRAS.482.5477W} {482, 5477}

\bibitem[\protect\citeauthoryear{{Weinzirl}, {Jogee}, {Khochfar}, {Burkert}  \&
  {Kormendy}}{{Weinzirl} et~al.}{2009}]{WJK09}
{Weinzirl} T.,  {Jogee} S.,  {Khochfar} S.,  {Burkert} A.,   {Kormendy} J.,
  2009, \mn@doi [\apj] {10.1088/0004-637X/696/1/411}, \href
  {https://ui.adsabs.harvard.edu/abs/2009ApJ...696..411W} {696, 411}

\bibitem[\protect\citeauthoryear{{Willett} et~al.,}{{Willett}
  et~al.}{2013}]{WLB13}
{Willett} K.~W.,  et~al., 2013, \mn@doi [\mnras] {10.1093/mnras/stt1458}, \href
  {https://ui.adsabs.harvard.edu/abs/2013MNRAS.435.2835W} {435, 2835}

\bibitem[\protect\citeauthoryear{{Zanisi} et~al.,}{{Zanisi}
  et~al.}{2020}]{ZSL20}
{Zanisi} L.,  et~al., 2020, \mn@doi [\mnras] {10.1093/mnras/stz3516}, \href
  {https://ui.adsabs.harvard.edu/abs/2020MNRAS.492.1671Z} {492, 1671}

\bibitem[\protect\citeauthoryear{{Zavala}, {Avila-Reese},
  {Hern{\'a}ndez-Toledo}  \& {Firmani}}{{Zavala} et~al.}{2003}]{ZAH03}
{Zavala} J.,  {Avila-Reese} V.,  {Hern{\'a}ndez-Toledo} H.,   {Firmani} C.,
  2003, \mn@doi [\aap] {10.1051/0004-6361:20031135}, \href
  {https://ui.adsabs.harvard.edu/abs/2003A&A...412..633Z} {412, 633}

\bibitem[\protect\citeauthoryear{{Zhu} et~al.,}{{Zhu} et~al.}{2018}]{ZVV18}
{Zhu} L.,  et~al., 2018, \mn@doi [Nature Astronomy]
  {10.1038/s41550-017-0348-1}, \href
  {https://ui.adsabs.harvard.edu/abs/2018NatAs...2..233Z} {2, 233}

\bibitem[\protect\citeauthoryear{{Zonca}, {Singer}, {Lenz}, {Reinecke},
  {Rosset}, {Hivon}  \& {Gorski}}{{Zonca} et~al.}{2019}]{ZSL19}
{Zonca} A.,  {Singer} L.,  {Lenz} D.,  {Reinecke} M.,  {Rosset} C.,  {Hivon}
  E.,   {Gorski} K.,  2019, \mn@doi [The Journal of Open Source Software]
  {10.21105/joss.01298}, \href
  {https://ui.adsabs.harvard.edu/abs/2019JOSS....4.1298Z} {4, 1298}

\bibitem[\protect\citeauthoryear{{de Souza}, {Gadotti}  \& {dos Anjos}}{{de
  Souza} et~al.}{2004}]{SGA04}
{de Souza} R.~E.,  {Gadotti} D.~A.,   {dos Anjos} S.,  2004, \mn@doi [\apjs]
  {10.1086/421554}, \href
  {https://ui.adsabs.harvard.edu/abs/2004ApJS..153..411D} {153, 411}

\bibitem[\protect\citeauthoryear{{van der Hulst}, {van Albada}  \&
  {Sancisi}}{{van der Hulst} et~al.}{2001}]{VVS01}
{van der Hulst} J.~M.,  {van Albada} T.~S.,   {Sancisi} R.,  2001, in {Hibbard}
  J.~E.,  {Rupen} M.,   {van Gorkom} J.~H.,  eds,  Astronomical Society of the
  Pacific Conference Series Vol. 240, Gas and Galaxy Evolution. p.~451

\bibitem[\protect\citeauthoryear{{van der Walt}, {Colbert}  \&
  {Varoquaux}}{{van der Walt} et~al.}{2011}]{VCG11}
{van der Walt} S.,  {Colbert} S.~C.,   {Varoquaux} G.,  2011, \mn@doi
  [Computing in Science and Engineering] {10.1109/MCSE.2011.37}, \href
  {https://ui.adsabs.harvard.edu/abs/2011CSE....13b..22V} {13, 22}

\makeatother
\end{thebibliography}

\appendix
\section{Rotation matrix} \label{app:A}
We rotate the coordinate and velocity vectors of each galaxy's stellar particle by applying to each vector the dot product of following rotation matrices:
\begin{equation}
\textbf{R}_{z} = 
   \begin{vmatrix} 
   \mathrm{cos(\alpha)} & \mathrm{sin(\alpha)} & 0  \\
   \mathrm{-sin(\alpha)} & \mathrm{cos(\alpha)} & 0  \\
   0 & 0 & 1  \\
   \end{vmatrix} 
\end{equation}
and
	\begin{equation}
\textbf{R}_{y} = 
   \begin{vmatrix} 
   \mathrm{cos(\delta)} & 0 & \mathrm{sin(\delta)}  \\
   0 & 1 & 0  \\
   \mathrm{-sin(\delta)} & 0 & \mathrm{cos(\delta)}  \\
   \end{vmatrix} \; ,
\end{equation}
where $\alpha$ and $\delta$ are the right ascension and elevation from the reference plane of the galactic angular momentum.

\section{Profiles} \label{app:B}
The exact value for the term $b_n$ can be obtained by solving:
\begin{align} \label{eq:DeltaJgas}
\Gamma(2n) = 2\gamma(2n,b)\; ,
\end{align}
where $\Gamma$ and $\gamma$ are the complete and incomplete gamma functions, respectively \cite{C91}. Useful apporximations have been proposed by \cite{PS97,CB99}, however in this work w follow \cite{MCH03} and use the asymptotic expansion of \cite{CB99} for all $n>0.36$:
\begin{align} \label{eq:DeltaJgas}
\nonumber b_{n} &=\, 2n -\frac{1}{3} + \frac{4}{405n} + \frac{46}{25515n^{2}} + \frac{131}{1148175n^{3}} - \frac{2194697}{30690717750n^{4}} \\
&+ O(n^{-5})\; ,
\end{align}
and the polynomial expression derived by \cite{MCH03} for $n \leq 0.36$:
\begin{align} \label{eq:DeltaJgas}
b_{n} = \sum_{i=0} ^{m}\alpha_{i} n^{i}\; ,
\end{align}
where $m$ is the order of the polynomial and $\alpha_{i}$ are the coefficients of the fit which can be written as:
\begin{align} \label{eq:DeltaJgas}
\alpha_{0} &= 0.01945\; \alpha_{1} = -0.8902\; ,\nonumber\\ 
\alpha_{2} = 10&.95\; \alpha_{3} = -19.67\; , \alpha_{4} = 13.43\; .
\end{align}

\label{lastpage}
\end{document}